\pdfoutput=1
\newcommand{\DOCCLASS}{acmsmall}
\def\ACMPAPER{}

\documentclass[\DOCCLASS]{acmart}
\usepackage{booktabs} 
\usepackage[english]{babel}
\usepackage{graphicx}
\usepackage{amsmath}
\usepackage{cancel}
\usepackage{amsfonts}
\usepackage{amssymb}
\usepackage[noend]{algpseudocode}
\usepackage{listingsutf8}
\usepackage{textcomp}
\usepackage{relsize}
\usepackage{hyperref}
\usepackage{xcolor}
\usepackage{amsthm}
\usepackage{tabularx}
\usepackage{mathrsfs}
\usepackage{csvsimple}
\usepackage{tikz}
\usetikzlibrary{backgrounds,calc,positioning,arrows.meta}

\usepackage{todonotes}
\usepackage{varwidth}

\errorcontextlines\maxdimen

\makeatletter
    \newcommand*{\algrule}[1][\algorithmicindent]{\makebox[#1][l]{\hspace*{.5em}\thealgruleextra\vrule height \thealgruleheight depth \thealgruledepth}}%
\newcommand*{\thealgruleextra}{}
\newcommand*{\thealgruleheight}{.75\baselineskip}
\newcommand*{\thealgruledepth}{.25\baselineskip}

\newcount\ALG@printindent@tempcnta
\def\ALG@printindent{%
    \ifnum \theALG@nested>0
        \ifx\ALG@text\ALG@x@notext
        \else
            \unskip
            \addvspace{-1pt}
            \ALG@printindent@tempcnta=1
            \loop
                \algrule[\csname ALG@ind@\the\ALG@printindent@tempcnta\endcsname]%
                \advance \ALG@printindent@tempcnta 1
            \ifnum \ALG@printindent@tempcnta<\numexpr\theALG@nested+1\relax
            \repeat
        \fi
    \fi
    }%
\usepackage{etoolbox}
\patchcmd{\ALG@doentity}{\noindent\hskip\ALG@tlm}{\ALG@printindent}{}{\errmessage{failed to patch}}

\renewcommand{\ALG@beginalgorithmic}{\small}
\makeatother

\newbox\statebox
\newcommand{\myState}[1]{%
    \setbox\statebox=\vbox{#1}%
    \edef\thealgruleheight{\dimexpr \the\ht\statebox+1pt\relax}%
    \edef\thealgruledepth{\dimexpr \the\dp\statebox+1pt\relax}%
    \ifdim\thealgruleheight<.75\baselineskip
        \def\thealgruleheight{\dimexpr .75\baselineskip+1pt\relax}%
    \fi
    \ifdim\thealgruledepth<.25\baselineskip
        \def\thealgruledepth{\dimexpr .25\baselineskip+1pt\relax}%
    \fi
    \State #1%
    \def\thealgruleheight{\dimexpr .75\baselineskip+1pt\relax}%
    \def\thealgruledepth{\dimexpr .25\baselineskip+1pt\relax}%
}

\usepackage[utf8]{inputenc}
\usepackage[T1]{fontenc}

\usepackage{algorithm}
\title{Interactive Runtime Verification}

\setcopyright{none}

\begin{document}

\acmYear{2017}
\acmMonth{5}



\author{Raphaël Jakse}
\affiliation{%
  \institution{Université Grenoble Alpes}
  \streetaddress{Antenne Inria Grenoble Giant - 17 rue des Martyrs - Bâtiment 50C 2ème étage
- Minatec Campus}
  \city{Grenoble}
  \postcode{38000}
}
\email{Raphael.Jakse@univ-grenoble-alpes.fr}

\author{Yliès Falcone}
\affiliation{%
  \institution{Université Grenoble Alpes}
  \streetaddress{Antenne Inria Grenoble Giant - 17 rue des Martyrs - Bâtiment 50C 2ème étage
- Minatec Campus}
  \city{Grenoble}
  \postcode{38000}
}
\email{Ylies.Falcone@univ-grenoble-alpes.fr}

\author{Jean-François Méhaut}
\affiliation{%
  \institution{Université Grenoble Alpes}
  \streetaddress{Antenne Inria Grenoble Giant - 17 rue des Martyrs - Bâtiment 50C 2ème étage
- Minatec Campus}
  \city{Grenoble}
  \postcode{38000}
}
\email{Jean-François.Mehaut@univ-grenoble-alpes.fr}

\author{Kevin Pouget}
\affiliation{%
  \institution{Université Grenoble Alpes}
  \streetaddress{Antenne Inria Grenoble Giant - 17 rue des Martyrs - Bâtiment 50C 2ème étage
- Minatec Campus}
  \city{Grenoble}
  \postcode{38000}
}
\email{Kevin.Pouget@univ-grenoble-alpes.fr}


%
%



\begin{abstract}
Monitoring is the study of a system at runtime, looking for input and output
events to discover, check or enforce behavioral properties. Interactive
debugging is the study of a system at runtime in order to discover and
understand its bugs and fix them, inspecting interactively its internal state.

Interactive Runtime Verification (i-RV) combines monitoring and interactive debugging.
We define an efficient and convenient way to check behavioral properties automatically on a program using a debugger.
We aim at helping bug discovery while keeping the classical debugging
techniques and interactivity, which allow understanding and fixing bugs.

\end{abstract}
\maketitle
\raggedbottom
\ifdefined \definition \else
\theoremstyle{definition}
\newtheorem{definition}{Definition}
\fi

\ifdefined \example \else
\theoremstyle{definition}
\newtheorem{example}{Example}
\fi

\ifdefined \proposition \else
\theoremstyle{remark}
\newtheorem{proposition}{Proposition}
\fi

\ifdefined \lemma \else
\theoremstyle{remark}
\newtheorem{lemma}{Lemma}
\fi

\ifdefined \theorem \else
\theoremstyle{remark}
\newtheorem{theorem}{Theorem}
\fi

\ifdefined \conditions \else
\theoremstyle{definition}
\newtheorem{condition}{Condition}
\fi

\theoremstyle{remark}
\newtheorem{remark}{Remark}

\theoremstyle{remark*}
\newtheorem{note*}{Note}

\theoremstyle{remark}
\newtheorem{assumption}{Assumption}

\newcommand{\Interactive}{     {\scriptstyle\mathrm{\mathbf{I}}}}
\newcommand{\Passive}{         {\scriptstyle\mathrm{\mathbf{P}}}}
\newcommand{\dbgcmd}[1]{       \mathtt{cmd} \; \mathrm{#1}}
\newcommand{\runInstr}{        \mathrm{runInstr}}

\ifdefined\TWOCOLS
	\newcommand{\instrumentnorm}{instr}
	\newcommand{\unInstrumentnorm}{unInstr}
	\newcommand{\ParamNames}{\mathrm{Params}}
	\newcommand{\wpEvtToAddressesnorm}{evtToWPs}
	\newcommand{\bpEvtToAddressnorm}{evtToBP}
	\newcommand{\applyEventsnorm}{applyEvts}
	\newcommand{\true}{\mathbf{T}}
	\newcommand{\false}{\mathbf{F}}
    \newcommand{\trueexplicit}{\true\,\textrm{(true)}}
    \newcommand{\falseexplicit}{\false\,\textrm{(false)}}
\else
	\newcommand{\instrumentnorm}{instrument}
	\newcommand{\unInstrumentnorm}{unInstrument}
	\newcommand{\ParamNames}{\mathrm{ParamNames}}
	\newcommand{\wpEvtToAddressesnorm}{evtToWatchpoints}
	\newcommand{\bpEvtToAddressnorm}{evtToBreakpoint}
	\newcommand{\applyEventsnorm}{applyEvents}
	\newcommand{\true}{\mathrm{true}}
	\newcommand{\false}{\mathrm{false}}
    \newcommand{\trueexplicit}{\true}
    \newcommand{\falseexplicit}{\false}
\fi

\newcommand{\wpEvtToAddresses}{\textsc{\wpEvtToAddressesnorm}}
\newcommand{\bpEvtToAddress}{\textsc{\bpEvtToAddressnorm}}
\newcommand{\normalStepnorm}{normalStep}

\newcommand{\instrument}{      \textsc{\instrumentnorm}}
\newcommand{\unInstrument}{      \textsc{\unInstrumentnorm}}

\newcommand{\setBP}{           \mathrm{setBP}}
\newcommand{\setWP}{           \mathrm{setWP}}
\newcommand{\interactiveStep}{ \textsc{interactiveStep}}
\newcommand{\unsetBP}{         \mathrm{unsetBP}}
\newcommand{\unsetWP}{         \mathrm{unsetWP}}
\newcommand{\setInstrBp}{      \mathrm{setInstrBp}}
\newcommand{\evtToAddress}{    \mathrm{evtToAddress}}
\newcommand{\bpToEvts}{        \mathrm{bpToEvts}}
\newcommand{\wpsToEvts}{        \mathrm{wpsToEvts}}
\newcommand{\needsBreakpoint}[1]{ \mathrm{needsBreakpoint}(#1)}
\newcommand{\needsWatchpoint}{ \mathrm{needsWatchpoint}}
\newcommand{\FunctionCall}{    \mathrm{FunctionCall}}
\newcommand{\ValueWrite}{      \mathrm{ValueWrite}}
\newcommand{\ValueRead}{       \mathrm{ValueRead}}
\newcommand{\UpdateExpr}{      \mathrm{UpdateExpr}}
\newcommand{\Sym}{             \mathrm{Sym}}
\newcommand{\name}{            \mathrm{name}}
\newcommand{\type}{            \mathrm{type}}
\newcommand{\params}{          \mathrm{params}}
\newcommand{\values}{          \mathrm{values}}
\newcommand{\enabled}{         \mathrm{enabled}}
\newcommand{\BREAK}{           \mathrm{\mathbf{B}}}
\newcommand{\isBefore}{        \mathrm{isBefore}}
\newcommand{\instr}{           \mathit{instr}}
\newcommand{\Instr}{           \mathrm{Instr}}
\newcommand{\init}{            \mathrm{init}}
\newcommand{\Env}{             \mathrm{Env}}
\newcommand{\Events}{          \mathrm{Events}}
\newcommand{\env}{             \mathrm{env}}
\newcommand{\addr}{            \mathit{addr}}
\newcommand{\dk}{              \mathrm{DK}}
\newcommand{\isBeforeEvt}{     \mathrm{isBeforeEvt}}
\newcommand{\beforeEvt}{       \mathrm{before}_{\mathrm{ev}}}
\newcommand{\afterEvt}{        \mathrm{after}_{\mathrm{ev}}}
\newcommand{\Mem}{             \mathrm{Mem}}
\newcommand{\Checkpoints}{     \mathcal C \mathrm{p}}
\newcommand{\Checkp}{          \mathrm{\mathscr C}}
\newcommand{\Breakpoints}{     \mathcal B \mathrm{p}}
\newcommand{\Bp}{              \mathrm{\mathscr B}}
\newcommand{\Watchpoints}{     \mathcal W \mathrm{p}}
\newcommand{\Wp}{              \mathrm{\mathscr W}}
\newcommand{\Address}{         \mathrm{Address}}
\newcommand{\InstrAddress}{    \mathrm{InstrAddress}}
\newcommand{\DBGInst}{         \mathrm{DBGInst}}
\newcommand{\isUserBreakpoint}{\mathit{isUserBP}}
\newcommand{\isUserWatchpoint}{\mathit{isUserWP}}
\newcommand{\isuserwp}{        \isUserWatchpoint}
\newcommand{\isuserbp}{        \isUserBreakpoint}
\newcommand{\st}{              \mathrm{\; \text{s.t.} \;}}
\newcommand{\pc}{              \mathit{pc}}
\newcommand{\stopInstr}{       \mathrm{\mathbf{stop}}}
\newcommand{\removeAllBPs}{    \mathrm{removeAllBPs}}
\newcommand{\restoreAllBPs}{   \mathrm{restoreBPs}}
\newcommand{\getValueInPrgm}{  \mathrm{getValueInPrgm}}
\newcommand{\getAddrParamN}{   \mathrm{getAddrParamN}}
\newcommand{\getParamN}{       \mathrm{getParamN}}
\newcommand{\Names}{           \mathrm{Names}}
\newcommand{\Values}{          \mathrm{Values}}
\newcommand{\userbp}{          \mathrm{USER\_BP}}
\newcommand{\internalbp}{      \mathrm{INTERNAL\_BP}}
\newcommand{\stepbystepbp}{    \mathrm{STEP\_BY\_STEP\_BP}}
\newcommand{\bptype}{          \mathrm{bpType}}
\newcommand{\ret}{\mathrm{ret}}
\newcommand{\applyScenario}{\lambda_{\textsc{sc}}}
\newcommand{\ScenarioReaction}{\mathrm{SR}}
\newcommand{\scenarioReactionMatchesEvent}{SRMatchesEvt}
\newcommand{\B}{\mathbb{B}}
\newcommand{\Bptype}{\mathrm{B_T}}
\newcommand{\watchread}{\mathit{read}}
\newcommand{\watchwrite}{\mathit{write}}
\newcommand{\EventTypes}{\mathrm{EventTypes}}
\newcommand{\accesses}{\mathrm{accesses}}

\newcommand{\When}{\mathrm{When}}
\newcommand{\What}{\mathrm{Point}}
\newcommand{\before}{\mathtt{before}}
\newcommand{\after}{\mathtt{after}}
\newcommand{\entering}{\mathtt{entering}}
\newcommand{\leaving}{\mathtt{leaving}}
\newcommand{\A}{\mathrm{Actions}}
\newcommand{\E}{\mathrm{Expr}}
\newcommand{\uninstantiated}{\cancel{\mathtt{v}}}

\newcommand{\applyEvents}{\textsc{\applyEventsnorm}}

\newcommand{\applyEvent}{\textsc{applyEvent}}
\newcommand{\popBPnorm}{popBP}
\newcommand{\popBP}{\textsc{\popBPnorm}}

\newcommand{\normalStep}{\textsc{\normalStepnorm}}

\newcommand{\wholeState}{\confIRV}

\newcommand{\interactiveStepnorm}{interactiveStep}

\newcommand{\handleBPnorm}{handleBP}
\newcommand{\handleBP}{\textsc{\handleBPnorm}}
\newcommand{\handleWPnorm}{handleWP}
\newcommand{\handleWP}{\textsc{\handleWPnorm}}

\newcommand{\handleStepBPnorm}{handleStepBP}
\newcommand{\handleStepBP}{\textsc{\handleStepBPnorm}}

\newcommand{\handleStepWPnorm}{handleStepWP}
\newcommand{\handleStepWP}{\textsc{\handleStepWPnorm}}

\newcommand{\start}{\mathrm{start}}
\algnewcommand\Let{\textbf{let}$\;$}
\algnewcommand\algorithmicswitch{\textbf{switch}}
\algnewcommand\algorithmiccase{\textbf{case}}
\algnewcommand\algorithmicassert{\texttt{assert}}
\algnewcommand\Assert[1]{\State \algorithmicassert(#1)}%
\algdef{SE}[SWITCH]{Switch}{EndSwitch}[1]{\algorithmicswitch\ #1\ \algorithmicdo}   {\algorithmicend\ \algorithmicswitch}%
\algdef{SE}[CASE]{Case}{EndCase}[1]{\algorithmiccase\ #1}{\algorithmicend\     \algorithmiccase}%
\algtext*{EndSwitch}
\algtext*{EndCase}%

\newcommand\fixme[1]{\textcolor{orange}{FIXME: #1}}

\providecommand{\getValueInPrgm}{\mathrm{getValueInPrgm}}

\mathchardef\breakingcomma\mathcode`\,
{\catcode`,=\active
	\gdef,{\breakingcomma\discretionary{}{}{}}
}
\newcommand{\inmathlist}[1]{\mathcode`\,=\string"8000 #1}
\newcommand{\mathlist}[1]{$\inmathlist{#1}$}

\newcommand{\hzero}[1]{\section{#1}}
\newcommand{\hone}[1]{\subsection{#1}}
\newcommand{\htwo}[1]{\subsubsection{#1}}

\algdef{SE}[DOWHILE]{Do}{DoWhile}{\algorithmicdo}[1]{\algorithmicwhile\ #1}%

\ifdefined \BLINDVERSION
\newcommand{\ce}{\textit{i-RV Tool}}
\newcommand{\Ce}{\textit{i-RV Tool}}
\else
\newcommand{\ce}{Verde}
\newcommand{\Ce}{Verde}
\fi

\newcommand{\afaire}[1]{\todo{\footnotesize #1}}

\newcommand{\secref}[1]{Sec.~\ref{#1}}
\newcommand{\defref}[1]{Def.~\ref{#1}}
\newcommand{\algoref}[1]{Alg.~\ref{#1}}
\newcommand{\exref}[1]{Ex.~\ref{#1}}
\newcommand{\theoref}[1]{Theo.~\ref{#1}}
\newcommand{\figref}[1]{Fig.~\ref{#1}}

\newcommand{\yf}[1]{\todo[inline]{#1}}

\definecolor{dkgreen}{rgb}{0,0.6,0}
\definecolor{gray}{rgb}{0.4,0.4,0.4}
\definecolor{lightgray}{rgb}{0.9,0.9,0.9}
\definecolor{mauve}{rgb}{0.58,0,0.82}
\lstdefinestyle{CheckExecStyle}{
basicstyle=\footnotesize,
language=Python,
deletekeywords={max},
keywords={state,transition,non,accepting,failure,success,event,init,initialization,slice on},
numbers=left,
stepnumber=1,
numbersep=10pt,
tabsize=4,
showspaces=false,
showstringspaces=false,
numberstyle=\tiny\color{gray},
keywordstyle=\color{blue},
commentstyle=\color{dkgreen},
stringstyle=\color{mauve},
frame=single,
columns=fullflexible
}

\lstdefinestyle{pseudoCodeStyle}{
basicstyle=\footnotesize,
deletekeywords={max},
keywords={if,then,else,do,before,after,entering,state},
numbers=left,
stepnumber=1,
numbersep=10pt,
tabsize=4,
showspaces=false,
showstringspaces=false,
numberstyle=\tiny\color{gray},
keywordstyle=\color{blue},
commentstyle=\color{dkgreen},
stringstyle=\color{mauve},
frame=single,
columns=fullflexible
}

\newcommand{\hideInShortVersion}[1]{\ifdefined \SHORTVERSION \else #1 \fi}
\newcommand{\hideInRLYShortVersion}[1]{\ifdefined \RLYSHORTVERSION \else #1 \fi}

\ifdefined \SHORTVERSION
\def\HIDESCENARIOFORMALISM{}
\newenvironment{itemizeopt}{}{}
\newenvironment{itemizesquishopt}{}{}
\newcommand{\itemopt}[1][]{\ifthenelse{\equal{#1}{!}}{\!}{\!#1}}
\newcommand{\optdot}{.}
\newcommand{\Commentopt}[1]{}
\else
\newenvironment{itemizeopt}{: \begin{itemize}}{\end{itemize}}
\newenvironment{itemizesquishopt}{: \squishlist}{\squishend}
\newcommand{\itemopt}[1][]{\item}
\newcommand{\optdot}{}
\newcommand{\Commentopt}[1]{\Comment{#1}}
\fi

\newcommand{\squishlist}{
 \begin{list}{-}
  { \setlength{\itemsep}{0pt}
     \setlength{\parsep}{1pt}
     \setlength{\topsep}{1pt}
     \setlength{\partopsep}{0pt}
     \setlength{\leftmargin}{0.6em}
        \setlength{\labelwidth}{1.5em}
     \setlength{\labelsep}{0.4em} } }
\newcommand{\squishend}{
  \end{list}  }

\newcommand\hide[1]{\phantom{\varwidth{\linewidth}#1\endvarwidth}}

\ifdefined \ACMPAPER
\newcommand{\myparag}[1]{\vspace{-0.5em}\paragraph{#1}}
\else
\newcommand{\myparag}[1]{\paragraph{#1}}
\fi

\newcommand{\irv}{\textrm{i-RV}}
\newcommand{\confIRV}{c_{\irv}}

\newcommand{\confIRVinit}{\confIRV^{\mathrm{init}}}

\newcommand{\F}{\mathrm{Upd}}
\newcommand{\correspTo}{\rightsquigarrow}
\newcommand{\instrOrigOrBPPredicate}{\textsc{ins\-tr\-Orig}}
\newcommand{\bpConsistentPredicate}{\textsc{bp\-Con\-sis\-tent}}
\newcommand{\irvConfCorrespToPrgmPredicate}{\textsc{irv\-Cor\-resp\-To\-Prgm}}
\newcommand{\R}[1]{$#1$ satisfies the three predicates}
\newcommand{\pointtwo}[1]{$#1$ satisfies point two of \theoref{theo}}

\newcommand{\add}[1]{{\textcolor{red}{#1}}}
\newcommand{\updatemonnorm}{updateMon}
\newcommand{\updateMon}{\textsc{\updatemonnorm}}
\newcommand{\dom}[1]{\mathcal{D}(#1)}

\newcommand{\zsh}{\texttt{zsh}}

\ifdefined\RLYSHORTVERSION
\newcommand{\ShortIf}[2]{\State\algorithmicif\ #1 #2}
\newcommand{\TernaryAss}[4]{
    \State #1 $\gets$ \algorithmicif \, #2: #3
    \State \hide{#1 $\gets$ } \algorithmicelse \, #4
}
\else
\newcommand{\ShortIf}[2]{\If{#1} \State #2 \EndIf}
\newcommand{\TernaryAss}[4]{\If{#2} \State #1 $\gets$ #3 \Else \State #1 $\gets$ #4 \EndIf}
\fi

\ifdefined\HIDESCENARIOFORMALISM
\newcommand{\commamsprime}{}
\newcommand{\commams}{}
\else
\newcommand{\commams}{, m_s}
\newcommand{\commamsprime}{\commams'}
\fi

%

\hzero{Introduction}
\label{chap:introduction}

When developing software, detecting and fixing bugs as early as possible is important.
This can be difficult: an error does not systematically lead to a crash, it can remain undetected during the development cycle.
Besides, when detected, a bug can be hard to understand, especially if the method of detection does not provide methods to study the bug.
\paragraph{Interactive debugging}
A widespread way to fixing bugs consists in observing a bad behavior and starting a debugging session to find the cause.
A debugging session generally consists in repeating the following steps: executing the program in a debugger, setting breakpoints before the expected cause of the bug, finding the point in the execution where it starts being erratic and inspecting the internal state~(callstack, values of variables) to determine the cause of the problem.
The program is seen as a white box and its execution as a sequence of program states that the developer inspects step by step using a debugger in order to understand the cause of a misbehavior.
The execution is seen at a low level~(assembly code, often mapped to the source code) while one would ideally want it be abstracted.
The debugger links the binary code to the programming language.
The state of the program can be modified at runtime: variables can be edited, functions can be called, the execution can be restored to a previous state.
This lets the developer test hypotheses on a bug without having to modify the code, recompile and rerun the whole program, which would be time consuming.
However, this process can be tedious and prone to a lot of trials and errors.
Moreover, observing a bug does not guarantee that this bug will appear during the debugging session, especially if the misbehavior is caused by a race condition or a special input that was not recorded when the bug was observed.
Interactive debugging does not target bug discovery: usually, a developer already knows the bug existence and tries to understand it.

\paragraph{Monitoring}
Runtime verification (aka monitoring)~\cite{DBLP:conf/vstte/HavelundG05,DBLP:journals/jlp/LeuckerS09,Sokolsky2012} aims at detecting bugs.
The execution is abstracted into a sequence of events of program-state updates.
Monitoring aims at detecting misbehaviors of a black-box system:
its internal behavior is not accessible and its internal state generally cannot be altered.
Information on the internal state can be retrieved by instrumenting the execution of the program.
The execution trace can be analyzed offline~(i.e. after the termination of the program) as well as online~(i.e. during the execution) and constitutes a convenient abstraction on which it is possible to express runtime properties.

We aim at easing \textit{bug discovery}, \textit{bug understanding} as well as their \textit{combination}.
We introduce Interactive Runtime Verification (\irv{}), a method that brings bug discovery and bug understanding together by combining \emph{interactive debugging} and \emph{monitoring}, augmenting debuggers with runtime verification techniques.
Using \irv{}, one can discover a bug and start getting insight on its cause at the same time.
\hideInRLYShortVersion{
\begin{figure}
\begin{center}
	\newcommand{\spaceboxx}{3cm}
\newcommand{\spaceboxy}{1.5cm}

\resizebox{0.9\linewidth}{!}{%
\begin{tikzpicture}[
block/.style={
draw,
fill=white,
rectangle,
minimum width={width("Program Execution")*1.20},
minimum height={height("User")*7}
}
]
\node[align=center, block]                               (user)      {User};
\node[align=center, block, below=\spaceboxy of user]     (debugger)  {Debugger};
\node[align=center, block, below=\spaceboxy of debugger] (program)   {Program Execution};
\node[align=center, block, right=\spaceboxx of debugger] (monitor)   {Monitor \\ Checks a user-defined property};
\node[align=center, block, above=\spaceboxy of monitor]  (scenario)  {Scenario \\ Applies user-defined reactions};

\draw[align=center, -{Stealth[scale=2.0]}] (user.south) to node [auto] {Controls} (debugger.north);
\draw[align=center, -{Stealth[scale=2.0]}] (debugger.south) to node [auto] {Affects \\ Instruments} (program.north);
\draw[align=center, -{Stealth[scale=2.0]}] ($(debugger.east) +(0,0.25)$) to node [auto] {Events} ($(monitor.west) +(0,0.25)$);
\draw[align=center, -{Stealth[scale=2.0]}] ($(monitor.west) -(0,0.25)$) to node [auto] {Asks for \\ instrumentation} ($(debugger.east) -(0,0.25)$);
\draw[align=center, -{Stealth[scale=2.0]}] (monitor.north) to node [auto] {Events} (scenario.south);
\draw[align=center, -{Stealth[scale=2.0]}] (scenario.south west) to node [auto] {Controls} (debugger.north east);

\end{tikzpicture}
}
\end{center}
\caption{Overview of the approach}
\label{fig:schema-overview}
\end{figure}
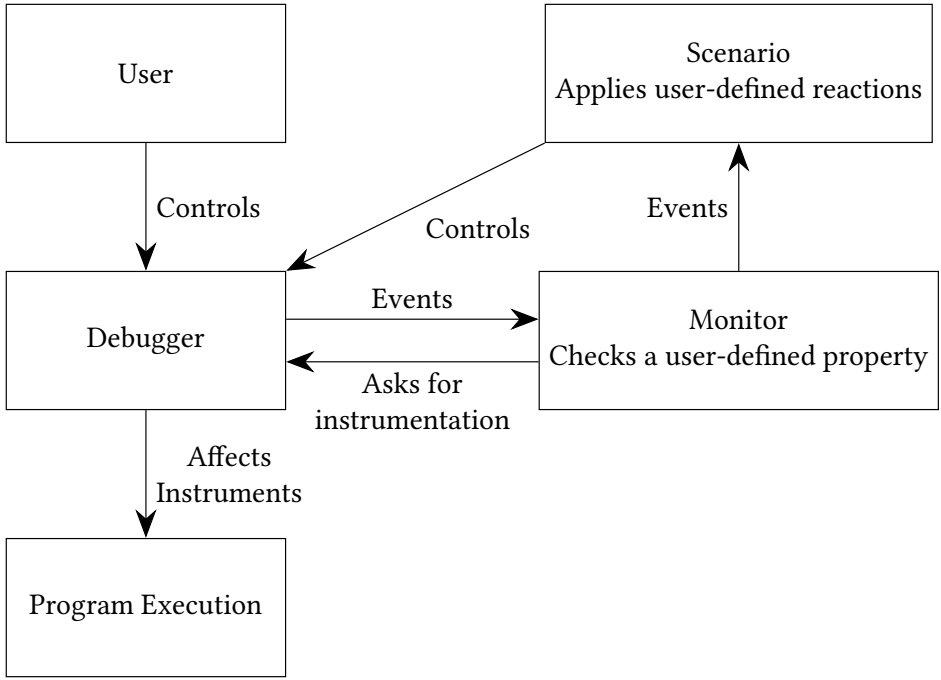
}
\irv{} aims at \textit{automating debugging}. For instance, it is possible to automatically stop the execution when a misbehavior is detected or to automate checkpointing at the right times.
We define an expressive \textit{property model} that allows flexibility when writing properties.
We give a \textit{formal description} of our execution model using high-level pseudo-code which serves as a \textit{basis for a solid implementation} and reasoning and to ensure correctness of our approach.
End-users are however not required to have a full understanding of this description.
\irv{} takes advantage of \textit{checkpoints}.
Checkpoints allow saving and restoring the program state.
They are a powerful tool to explore the behavior of programs by trying different execution paths.
\irv{} introduces the notion of \textit{Scenarios}.
They allow defining actions that are triggered depending on the current state of the property verification.
We provide a \textit{full-featured tool} for \irv{}, \ce{}, written in Python as a GDB extension, facilitating its integration to developers' traditional environment.
\Ce{} also provides an optional \textit{animated view} of the current state of the monitor.
We give a \textit{detailed evaluation} of \irv{} using \ce{}. This evaluation validates the usefulness of \irv{} and its applicability in terms of performance\hideInRLYShortVersion{.

\paragraph{Organization of this paper}
In \secref{sec:overview}, we give a general picture of our approach.
In \secref{chap:concepts}, we describe our approach more precisely.
In \secref{chap:implementation}, we present our proof-of-concept
implementation of this approach, \ce{}.
In \secref{sec:eval}, we evaluate our approach.
In \secref{sec:existing}, we present existing techniques for finding and
studying bugs and compare them to our work.
In \secref{chap:future}, we conclude by presenting our future works}\ifdefined\SHORTVERSION\footnote{An extended version of the paper is available at \url{https://gitlab.inria.fr/monitoring/verde}}\fi.
\hzero{Approach Overview}
\label{sec:overview}

\begin{figure}[t]
	\ifdefined \TWOCOLS
		\def\propfigsize{0.9}
	\else
		\def\propfigsize{0.5}
	\fi
	\centerline{\includegraphics[width=\propfigsize\linewidth]{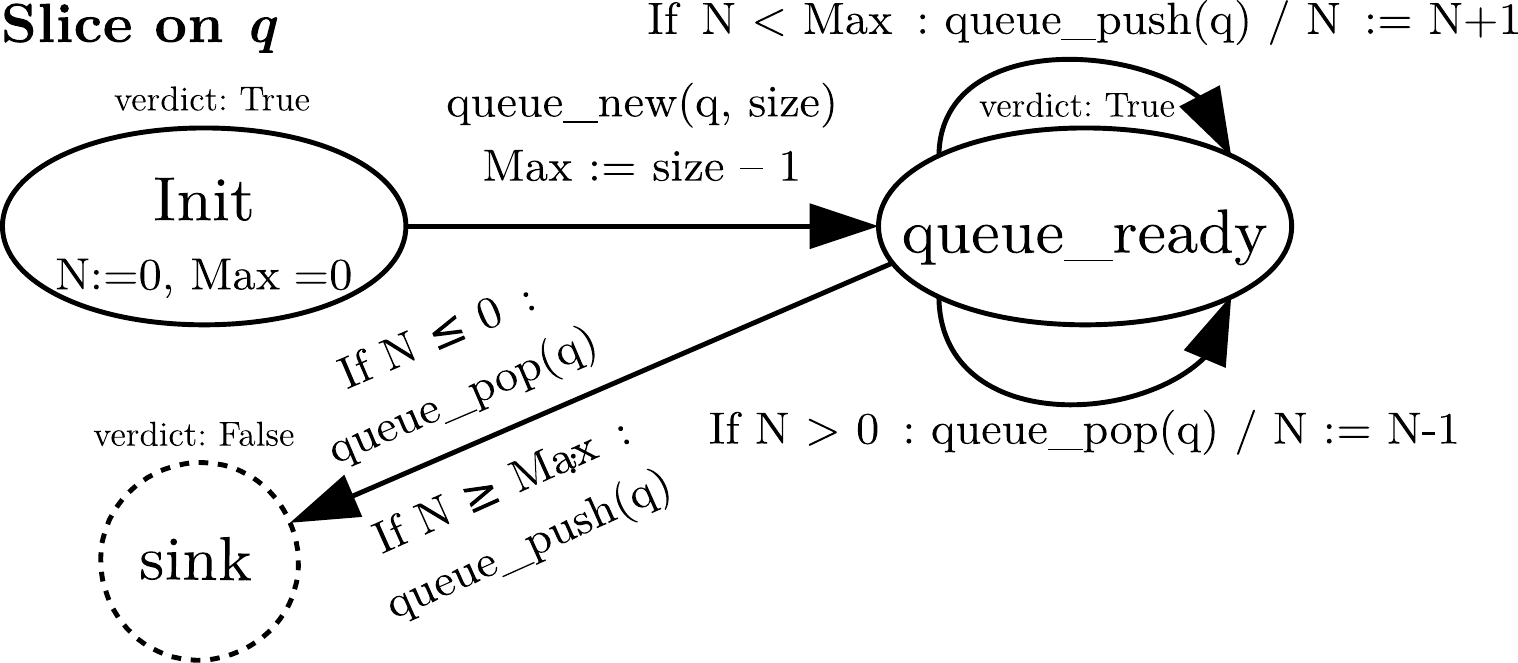}}
	\caption{Property for the absence of queue overflow in a producer-consumer program described
		in \secref{subsec:evalprodcons}.}
	\label{fig:exautomaton}
\end{figure}

In \irv{}\hideInRLYShortVersion{ (\figref{fig:schema-overview})},
the developer provides a property to check against the execution trace of a program to debug.
The property can be written according to its specification or the Application Programming Interface (API) of the libraries it uses.
An example of a property is pictured in Figure~\ref{fig:exautomaton} and gives the verdict false as soon as a queue overflows.
The program is run with a debugger which provides tools to instrument its execution, mainly breakpoints and watchpoints, and let us generate events to build the trace, including function calls and variable accesses.
An extension of the debugger provides a monitor that checks this property in real time.
Breakpoints and watchpoints are automatically set at relevant locations as the evaluation of a property requires monitoring function calls and memory accesses.
When an event stops influencing the evaluation of any property, the corresponding instrumentation (breakpoints, watchpoints) becomes useless and is therefore removed: the instrumentation is \emph{dynamic}.
The user-provided scenario defines what actions should be taken
during the execution according to the evaluation of the property.
Examples of scenarios are: when the verdict given by the monitor becomes
false (e.g. when the queue overflows), the execution is suspended to let the developer inspect and debug the program in the usual way, interactively; save the current state of the program (e.g. using a checkpoint, a feature provided by the debugger) while the property holds (e.g. while the queue has not overflown) and restore this state later, when the property does not hold anymore (e.g. at the moment the queue overflows).
When an event is generated --- when a breakpoint or a watchpoint is reached --- at  runtime, the monitor updates its state.
Monitor updates are seen as input events for the scenario.
Examples of these events are ``the monitor enters state X'',
``the state X has been left'', ``an accepting state has been entered'',
``a non-accepting state has been left''.

\hzero{Joint Execution of the Debugger, the Monitor and the Program}
\label{chap:concepts}
\label{chap:formal}

i-RV relies on the joint execution of different components: the program,
the debugger, the monitor and the scenario\ifdefined\HIDESCENARIOFORMALISM\footnote{Formal description of the scenario is given in the extended version of the paper.}\fi.
We formally describe the Interactively Runtime Verified program (\irv-program) composed of these components
as a Labeled Transition System (LTS). We first present each component
and our property model based on an extension of finite-state machines
in \secref{sec:components}.
Events play the role of symbols of the LTS.
Events are defined in \secref{sec:events}.
We then describe the evolution of the \irv-program in \secref{sec:evolconfigurations} using pseudo-code.
This formalization is not needed to adopt the approach.
However, it offers a solid basis for implementation and for reasoning and expressing properties over the concepts behind \irv. \ifdefined\SHORTVERSION In the extended version of this paper, we prove that the execution of the program is not affected by the presence of the monitor and the debugger and thus that all executions observed through i-RV are faithful.
\fi

\paragraph{Notations}
We define some notations used in this paper. We denote the set of booleans by $\B=\{\trueexplicit,\falseexplicit\}$
Given two sets $ E $ and $ F $, $ E \rightarrow F $ denotes the set of functions from $ E $ to $ F $.
By $f : E \rightarrow F $ or $f \in [ E \rightarrow F ]$, we denote that $f \in E \rightarrow F$ .
Let $f : E \longrightarrow F $, function $f' = f[x_1 \mapsto v]$ is such that $f'(x) = f(x)$ for any $x \neq x_1$, and $f'(x_1) = v$.
The domain of function $f$ is denoted by $\dom{f}$.

Let us consider a non-empty set of elements $E$.
The powerset of $E$ is denoted $\mathcal{P}(E)$.
Moreover, $\epsilon_E$ is the empty sequence (over $E$), noted $\epsilon$ when clear from the context.
$E^*$ denotes the set of finite sequences over $E$.
Given two sequences $ s $ and $ s' $, the sequence obtained by concatenating $s'$ to $s$ is denoted $s \cdot s'$.
We denote by $\Names$ the set of valid function and variable names in a program. 

We define the transitive relation ``\textit{$f'$ is more specific than $f$}'': $ f \sqsubseteq f' \overset{\mathrm{def}}{=} \dom{f} \subseteq \dom{f'} \land \forall p \in \dom{f}: f(p) = f'(p)$.
Likewise, ``\textit{$f'$ is strictly more specific than $f$}'': $ f \sqsubset f' \overset{\mathrm{def}}{=} \dom{f} \subset \dom{f'} \land \forall p \in \dom{f} \land f(p) = f'(p)$.

\hone{Events}
\label{sec:events}
%
i-RV is based on capturing events from the program execution with the debugger.
\hideInRLYShortVersion{
\begin{figure}
	\[
		\begin{array}{lcll}
			  p &  ::=  &  v  & (\textit{\footnotesize{a defined variable}})\\
						&   |   &  *p & (\textit{\footnotesize{the value pointed by p}}) \\
						&   |   &  \&p & (\textit{\footnotesize{the address of variable p}}) \\
						&   |   &  \arg i, i \in \mathbb{N} & (\textit{\footnotesize{the current value of parameter $i$}}) \\
			    &   |   &  \ret &
		\end{array}
	\]

	\caption{Grammar of valid parameter names}
	\label{fig:grammparams}
\end{figure}
}

\begin{definition}[Event]
	\label{def:event}
	An event is a tuple $e = (\inmathlist{t, n, p, i, b}) \in
	\EventTypes \times \Names \times \ParamNames^* \times \Values_p^* \times \B)$
	where $\EventTypes =
	\{\inmathlist{\mathtt{Call}, \mathtt{ValueWrite}, \mathtt{ValueRead}, \mathtt{UpdateExpr}}\}$.
	The event name $n \in \Names$ is denoted $\name(e)$.
	\ifdefined \RLYSHORTVERSION
	Valid parameter names in $\ParamNames$ are: $v$ (a defined variable), $*p$ (the value pointed by $p$, with $p \in \ParamNames$), $\&p$ (the address of variable $p$), $\arg i$ (the current value of parameter of index $i$) and $\ret$ (the ``return value'', which depends on the event type).
	\else
	A grammar describing the set of valid parameter names $\ParamNames$ is given in Figure~\ref{fig:grammparams}. A parameter can be the name of a variable defined in the program, the value pointed by a pointer, the address of a variable, an argument of the current function or a return value.
	\fi
 	\hideInRLYShortVersion{
	\begin{remark}
		The parameter $\arg i$ is not necessarily the value that was passed to the function when it was called.
		The parameter can be modified between the function call and when the event is triggered.
	\end{remark}
 	}
	If $e$ is a symbolic event, its parameters are uninstantiated, i.e.,
	$i=\emptyset$. If $e$ is a runtime event, $i$ is a list of parameter instances and
	\ifdefined \SHORTVERSION
		$\values(e) \colon \Names \rightarrow \Values$ maps parameters to their values:
		$(\values(e))(p_k) = i_k$.
	\else
		$\values(e)$ is the function that maps parameters to their values:
		\begin{align*}
		\values(e) \colon \Names & \; \longrightarrow \; \Values \\
		p_k	& \; \longmapsto	 \; i_k
		\end{align*}
	\fi
	Symbolic events are used to describe properties.
Runtime event are \textit{matched with} symbolic events if all its components, except $\values$, are identical to the components of the symbolic event.
\end{definition}
\begin{example}[Event]
	$(\FunctionCall, \mathtt{push}, (q, v), \emptyset, \true)$ is an event that is triggered before the call to function $\mathtt{push}$. Parameters $q$ and $v$ are retrieved when producing the event.
\end{example}
\hideInRLYShortVersion{
\begin{definition}[Event matching]
	A runtime event $e_i$ \textit{matches} a symbolic event $e_f$ if
	$\name(e_i)=\name(e_f)$ and $\type(e_i)=\type(e_f)$ and
	$\isBefore(e_i)=\isBefore(e_f)$ and $\params(e_i)=\params(e_f)$.
\end{definition}

\begin{example}[Event matching]
	``Before push(q, 5)'' is a runtime event matching the
	symbolic event ``Before a call to function (type) push (name) that
	takes a queue and an element in parameters (list of parameters)''.
\end{example}
}
The type $t \in \EventTypes$ of event $e$ is denoted $\type(e)$.
If $ b = \true$ (resp. $\false$), $e$ is a before (resp. after) event and $ \isBefore(e) = \true $ (resp. $\false$).
We describe the different event types.
A $\FunctionCall$ event is generated by a function call.
A before event is triggered before the first instruction of the function and after the jump to the function body. An after event fires after the last instruction of the function and before the jump to the caller.
The parameter $\ret$ then corresponds to the return value of the call.
A $\ValueWrite$ event is generated by an assignment.
A before (resp. after) event fires before (resp. after) the assignment instruction and parameter $\ret$ refers to the old (resp. new) value of the variable. A $\ValueRead$ event is generated by a variable read.
A before event fires before (resp. after) the instruction that reads the variable and parameter $\ret$ refers to the value of the variable.
An $\UpdateExpr$ event is generated whenever the value of an expression is changed.
A before (resp. after) event $e$ is fired before (resp. after) the update.
For a before (resp. after) $\UpdateExpr$ event, parameter $\ret$ refers to the old (resp. new) value of the expression.

\hideInRLYShortVersion{
\begin{remark}
	In practice, $\FunctionCall$ events are captured using breakpoints and
	$\ValueWrite$, $\ValueRead$ and $\UpdateExpr$ events are
	captured using watchpoints. An $\UpdateExpr$ event requires as
	many watchpoints as variables in the expression. Current
	debuggers hide this requirement by allowing setting watchpoints on
	expressions.
\end{remark}
}

%
\hone{Modeling the Components of i-RV}
\label{sec:components}
%
We model the components of i-RV and their behaviors by giving their configurations.
Our execution model is a composition of these configurations.

%
\htwo{The Program}
\label{sec:prgm}
%
For the sake of generality, we define a platform-independent and language-independent abstraction of a program that is loaded in memory, which allows us to apply the runtime techniques used in i-RV.
The memory is abstracted as a function that maps addresses to values.

\begin{definition}[Memory]
	A memory $ m $ is a function in $ \Mem = [\Address \rightarrow \Values] $. Some addresses correspond to variables of the program and are therefore linked to symbol names by the \textit{symbol table} built during the compilation of the program.
\end{definition}
\hideInRLYShortVersion{
\begin{remark}
The actual type of the elements of  $\Address$ does not matter.
They can be seen as integers like in a real memory.
Elements of $\Values$ are machine words.
They are either data (values of variables) or program instructions.
They can also be seen as integers.
\end{remark}
}

\begin{definition}[Program]
	\label{def:prgm}
	A \textit{program} is a 4-tuple $ (\inmathlist{\Sym, m_p^0 , \start, \runInstr})$ where\hideInShortVersion{:}\begin{itemizeopt}
		\itemopt[!]
			$\Sym : \Names \rightarrow \Address $ is a \emph{symbol table}
		\itemopt[!]
            $m_p^0 \in \Mem$ is the \emph{initial memory},
		\itemopt[!]
            $\start \in \Address$ is an address that points to the first instruction to run in the memory, and
		\itemopt[!]
            $\runInstr : (\Mem \times \Address) \rightarrow (\Mem \times \Address \times (\Address \times \B \times \B)^*) $ is a function that abstracts the operational semantics of the program\footnote{The actual semantics usually depends on the instruction set of the architecture.}.
	\end{itemizeopt}
\end{definition}

Function $\runInstr$ takes the current memory and Program Counter (PC) (in $\Address$) and executes the instruction at PC: it returns a (possibly new) memory, a new PC and a list of 3-tuples made of an address, and two booleans, representing the accesses to the memory.
In an access, the two booleans hold true if the value at the given address was read and written~(respectively), false otherwise.
Memory accesses are used by the debugger to trigger watchpoints~(see \secref{sec:debugger}).

\begin{example}[Program]
\label{ex:program}
	In the remainder of this section, we will use program $P$ given by the following source code to illustrate the concepts:
\begin{lstlisting}
a := 0 ; b := 1 ; a := a + b
\end{lstlisting}
\end{example}

\begin{definition}[Configuration of the program]
\label{def:conf_prog}
A \textit{configuration} is a pair $ (m_p, \pc) \in \Mem \times \Address $ where $m_p$ is the memory and $\pc$ is the current PC~(an address in the program memory), i.e. the address of the next instruction.
\end{definition}

\begin{example}[Configuration of the program]
    For program $P$ given in Ex.~\ref{ex:program}, just after the execution of the second instruction, the configuration of the program is $(m_p, \pc_3)$ where $\pc_3$ is the address of the code that corresponds to the third instruction of $P$, $m_p[\Sym(a)] = 0]$ and $m_p[\Sym(b)] = 1]$.
\end{example}
%
\htwo{The Monitor}
\label{sec:model}
%
The monitor evaluates a property against a trace, giving a verdict upon the reception of each event.
The verdict corresponding to the last event of the execution trace is called the final verdict~\cite{DBLP:series/natosec/FalconeHR13}.

\myparag{Property model}
We describe properties in a model based on finite-state machines.
It is composed of states, transitions and an environment and it recognizes sequences of events.
Transitions have guards that are expressions of event parameters and the memory and a function that can update the environment.
Properties can be expressed on the whole set of events that can be retrieved from the debugger.
Events are parameterized, i.e. values are linked to events.
For instance, a function call generates an event parameterized with the values of arguments passed during this call, as well as values that are accessible at this time (global variables for example).

\myparag{Trace slicing}
Some properties should hold on each instance of an object or a set of objects rather than on the global state of the program.
For example, a property on good file usage must be checked on each file that is manipulated by the program.
For these properties, the execution trace is sliced in a way that is similar to what is achieved by trace slicing in~\cite{DBLP:conf/tacas/ChenR09,DBLP:conf/syde/HavelundR15}.
Each slice of the trace concerns a specific instance of an object or a set of objects on which the property holds.
When trace slicing is used, a monitor does not correspond to a single finite state machine but to a set of finite state machines, one for each particular instance of an object.

\begin{definition}[Monitor]
	\label{def:monitor}

	A \textit{monitor} is a 7-tuple $(\inmathlist{Q, \Sigma, \init, \env_0, \Delta, v, S})$ where
	$Q$ is a set of states, $\Sigma$ is the set of symbolic events, $\env_0 \in \Env$
	is the initial environment~($\Env=\Names \rightarrow \Values$, where $\Names$
	is the set of variable names and $\Values$ is the set of values that can be
	stored in a variable), $ \Delta: \mathcal{P}(\Names \times \Names) \times Q
	\times \Sigma \times (\Env \times \Env
	\rightarrow \B) \times (\Env \times \Env \rightarrow \Env) \times Q $ is the
	transition relation,  $v \in [Q \rightarrow \mathcal{V}]$ is the function
	that maps states to verdicts and $S \subseteq \Names$ is a set of parameter
	names on which the slicing applies.

	A \textit{transition} is a 6-tuple $(sb, q_s, e_f, g, upd, q_d)$ where
	$sb$ is the slice binding of the transition,
	$q_s$ is the start state,
	$e_f$ is the symbolic event,
	$g$ is the guard,
	$upd$ is the ``updater'' and
	$q_d$ is the destination state.
	The slice binding $sb$ is a set of pairs $(p, s)$ where $p$ is a name of the parameter of the function on which the
	slicing applies and $s \in S$ is the name of the slice parameter at the level of the property.
	\ifdefined \RLYSHORTVERSION \else
	\begin{remark}
		In practice, in most cases, $p$ and $s$ are equal. However, it is possible that a particular object is named differently in different functions. This is the reason why a slice binding is used instead of a simple parameter name: an object is uniquely identified in $S$ (at the level of the property), and then each actual function parameter name is bound to this unique identifier.
	\end{remark}
	\fi
	The guard $g : \Env \times \Env \rightarrow \B $ takes the environment built from the parameters of the runtime event,
	the environment of the monitor and returns a boolean.
	If it returns true (resp. false), the transition is taken (resp. not taken).
	The updater $upd: \Env \times \Env \rightarrow \Env $ returns an environment from the environment built from the parameters of the runtime event and the environment of the monitor.
	This function is used to update the environment of the property.
\end{definition}

\hideInRLYShortVersion{
A monitor is given in \exref{ex:formalprop}.
}
\newcommand{\Max}{\mathrm{Max}}
\newcommand{\BBB}{\ifdefined\TWOCOLS \\ \quad \fi}

\begin{example}[Monitor]
	\label{ex:formalprop}
	The property illustrated in \figref{fig:exautomaton} is a tuple $(Q, \Sigma, \init, \env_0, \Delta, v, S)$ where:
	\begin{itemizesquishopt}
		\itemopt[!]
		$Q=\{\mathrm{Init}, \mathrm{ready}, \mathrm{sink}\}$,
		\itemopt[!]
		$\Sigma = \{\inmathlist{e_{f}^{\before}(\mathrm{queue\_new}), e_{f}^{\before}(\mathrm{push}), e_{f}^{\before}(\mathrm{pop})}\}$,
		\itemopt[!]
		$\init=\mathrm{Init}$,
		\itemopt[!]
		$\env_0=[N \mapsto 0, \Max \mapsto 0]$,
		\itemopt[!]
		$v=[\mathrm{Init} \mapsto \true, \mathrm{ready}=\true, \mathrm{sink}=\false]$,
		\itemopt[!]
		$S=\{q\}$,
		\itemopt[and]
		the transition $ \Delta $ is defined as $ \Delta = \{ $%
		{\small \vspace{-0.7em} \[
            \begin{array}{l}
                \quad (\{(q,q)\}, \mathrm{Init}, e_{f}^{\before}(\mathrm{new}), [any \mapsto \true],\BBB{} ([\mathrm{size}], \env) \mapsto \env[max := size - 1], \mathrm{ready}), \\
                \quad (\{(q,q)\}, \mathrm{ready}, e_{f}^{\before}(\mathrm{push}), [ [N, \Max] \mapsto N < \Max],\BBB{} (any, \env) \mapsto \env[N += 1], \mathrm{ready}), \\
                \quad (\{(q,q)\}, \mathrm{ready}, e_{f}^{\before}(\mathrm{pop}),  [ [N, \Max] \mapsto N > 0],\BBB{} (any, \env) \mapsto \env[N -= 1], \mathrm{ready}), \\
                \quad (\{(q,q)\}, \mathrm{ready}, e_{f}^{\before}(\mathrm{push}), [ [N, \Max] \mapsto N >= \Max],\BBB{} (any, \env) \mapsto \env, \mathrm{sink}), \\
                \quad (\{(q,q)\}, \mathrm{ready}, e_{f}^{\before}(\mathrm{pop}),  [ [N, \Max] \mapsto N <= 0],\BBB{} (any, \env) \mapsto \env, \mathrm{sink}) \\
                {\normalsize\}}
            \end{array}
        \]}
        The first transition makes the property transition from Init to ready when queue\_new is called.
		The guard always returns true so the transition is taken unconditionally.
		The updater stores the maximum number of elements in the queue in the environment of the monitor.
		This maximum is computed from the size parameter of the event new.
		The two next transitions make the monitor stay on the state ready when it is correct to add or (resp. remove) elements from the queue. In each case, the updater updates the number of elements in the queue in the environment of the monitor.
		The two last transitions detect that an element is added (resp. removed) though the queue is full (resp. empty) and makes the property transition from ready to sink.
		Each time a new value of the parameter $q$ is encountered, a new instance of
		the property is created.
	\end{itemizesquishopt}
\end{example}

We define the configuration of monitors.

\begin{definition}[Configuration of the monitor]
	\label{def:configmonitor}
	A \textit{configuration of the monitor} is
	a set of 4-tuples $ M = \{\inmathlist{(q_m^0, m_m^0, s_m^0, sp^0_m), \dots (q_m^n, m_m^n, s_m^n, sp^n_m)}\} \in \mathcal{P}(Q_m
	\times (\Names \rightarrow \Values) \times (S \rightarrow \Values
	\cup \{\uninstantiated\})\times (S \rightarrow \Values
	\cup \{\uninstantiated\}))$
	where $\uninstantiated$ corresponds to an uninstantiated value.
\end{definition}

In a configuration of a monitor, each 4-tuple $(\inmathlist{q_m^k, m_m^k, s_m^k, sp_m^{k}}) \in M $ represents an instance of the extended automaton that corresponds to a slice of the trace. $q_m^k$ is its current state, $m_m^k$ its current environment, $s_m^k$ a mapping that gives which instance of the parameters this slice corresponds to (the slice instance) and $sp_m^k$ the parent slice instance of this slice, that is, the slice instance of the slice $sp$ from which this slice was created (because an event with parameters more specific than the parameter instance of $sp$ happened).
We denote by $ C_m $ the set of configurations of a monitor and by $\enabled(M)$ the set of symbolic events to which the monitor is ``sensitive'' in M:
For all $q$ in $Q_m$, $\enabled(q)$ can be determined statically:
$\enabled(q) = \{e \in \Events \; | \; \exists (sb, g, upd, q_d): (sb, q, e, g, upd, q_d) \in \Delta_m\}$.
See Example~\ref{ex:enabled} for an illustration of $\enabled(q)$.
When a runtime event $e_i$ is triggered, a transition
$(\inmathlist{q_s, e_f, g, \textrm{upd}, q_d})$ is taken if the current state is $q_s$, $e_i$ matches $e_f$ and $g(e_i, m_p)=\true$, where $m_m$ is
the current environment.
If so, the memory and the state of the property are updated:
$m_m'=\mathrm{upd}(e_i, m_p)$, where $m_m'$ denotes the new environment and $q_d$ becomes the new state.

\begin{example}
	We denote by $e_{f}^{\before}(\phi(\mathrm{params}))$ the symbolic before event $(\inmathlist{\FunctionCall, \phi, \mathrm{params},  \isBefore})$ corresponding to a call to function~$\phi$.
	\label{ex:enabled}
	For the property of Figure~\ref{fig:exautomaton}\begin{itemizesquishopt}\itemopt[, ]
	    $\mathrm{enabled}(\mathrm{Init})=\{e_{f}^{\before}(\mathrm{queue\_new})\}$\hideInShortVersion{
		\item
		$\mathrm{enabled}(\mathrm{ready})=\{\inmathlist{e_{f}^{\before}(\mathrm{push}(q)),\ifdefined\TWOCOLS \\ \fi
		e_{f}^{\before}(\mathrm{pop}(q))}\}$
		\item
		$\mathrm{enabled}(\mathrm{sink})=\emptyset$}\optdot{ }
	\end{itemizesquishopt}
\end{example}

%
\htwo{The Debugger}
\label{sec:debugger}
%
The debugger provides primitives to instrument the program: breakpoints and watchpoints. It also provides a primitive to save the current state of the program and restore it: checkpoints.
These primitives can also be used by the user during an interactive debugging session. A breakpoint stops the execution at a given address $a \in Address$ and a watchpoint when a given address containing data of interest is accessed~(read, written, or both).

\begin{definition}[Breakpoint]
	\label{def:breakpoint}
	A \textit{breakpoint} is a 3-tuple~$(\inmathlist{\addr, \instr, \isuserbp}) $ where\begin{itemizeopt}
	\itemopt[!]
	$\addr \in \Address $ is the address of the breakpoint in the program memory,
	\itemopt[ ]
	$\instr \in \Values$ is the instruction to restore when the breakpoint is removed, and
	\itemopt[!]
	$\isuserbp \in \B$ is a boolean that holds $\true$ if the breakpoint was set by the user, and $\false$ if it was set by the monitor.
	\end{itemizeopt}
	The set of breakpoints is defined by $\Breakpoints = \Address \times \Instr \times \B$.
\end{definition}

\hideInRLYShortVersion{
As we shall see in \secref{sec:evolconfigurations}, when a breakpoint is reached, the execution is suspended and the debugger takes control over it.
When a breakpoint is set, the debugger stores the instruction that is at the address of the breakpoint to be able to restore it when the breakpoint is removed or when the instruction is to be executed.
}
\hideInRLYShortVersion{
Example~\ref{ex:breakpoint} illustrates the notion of breakpoint.
}
\begin{example}[Breakpoint]
\label{ex:breakpoint}
	A breakpoint set by the user on the second instruction of the program given in \exref{ex:program} is $(pc_2, b := 1, true)$ where $\pc_2$ is the memory address at which the second instruction is loaded.
	The instruction of the second instruction is stored as the second component of the tuple and the third component indicates that this breakpoint is set by the user.
\end{example}

%
\begin{definition}[Watchpoint]
	\label{def:watchpoint}
	A \textit{watchpoint} is a 4-tuple $(\inmathlist{\addr, \watchread, \watchwrite, \isuserwp}) \in \Watchpoints$ where\begin{itemizeopt}
		\itemopt[!]
			$\addr$ is the \emph{address} of the watchpoint in the program memory,
		\itemopt[!]
			$\watchread$ (resp. $\watchwrite$) is a Boolean that holds true if this watchpoint should be triggered when the memory is \emph{read} (resp. \emph{written}),
		\itemopt[!]
			$\isuserwp$ is a Boolean that holds $\true$ if the watchpoint was set by the user, and $\false$ if it was set by the monitor.\end{itemizeopt}
	The set of watchpoints is defined by $\Watchpoints = \Address \times \B \times \B \times \B$.
\end{definition}


\hideInRLYShortVersion{
Example~\ref{ex:watchpoint} illustrates the notion of watchpoint.
}
\begin{example}[Watchpoint]
\label{ex:watchpoint}
	A watchpoint set by the user on variable \texttt{b} in the program given in \exref{ex:program} is $(\&b, \false, \true)$ where $\&b$ denotes the address of variable \texttt{b} in the program memory.
	This watchpoint is triggered whenever variable \texttt{b} is written (but not when it is only read).
\end{example}

%
\paragraph{Checkpoint}
When debugging, it can be useful to save the state of the program (e.g., before the occurrence of a misbehavior to determine its cause or to try alternative executions).
A checkpoint can be set by the user as well as by the scenario. 
There is not syntactical element in the definition of a checkpoint as it only depends on runtime elements.
The states of the monitor and of the program are both saved, allowing coherent states after restoration.


%
\begin{definition}[Checkpoint]
	\label{def:checkpoint}
	A \textit{checkpoint} is a 2-tuple $(\inmathlist{c_p, c_m}) \in \Checkpoints$ where\begin{itemizeopt}
		\itemopt[!]
		$c_p$ is a configuration of the program (as per Definition~\ref{def:prgm}), and
		\itemopt[!]
		$c_m$ is a configuration of the monitor (as per Definition~\ref{def:monitor}).\end{itemizeopt}
	The set of all possible checkpoints is defined by $ \Checkpoints = (\Mem \times \Address) \times C_m $.
\end{definition}
%

%

\hideInRLYShortVersion{
Example~\ref{ex:checkpoint} illustrates the notion of checkpoint.
}
\begin{example}[Checkpoint]
	\label{ex:checkpoint}
	For the program given in \exref{ex:program}, the checkpoint $(([a \mapsto 0, b \mapsto 1], \pc_3), M)$, when the third instruction is about to be executed, is such that\ifdefined\SHORTVERSION\else:\fi
    \begin{itemizesquishopt}
        \itemopt[!]
            $[a \mapsto 0, b \mapsto 1]$ is the program memory,
        \itemopt[!]
            $\pc_3$ is the memory address at which the third instruction is loaded, and
        \itemopt[!]
            $M$ is the configuration of the monitor when the checkpoint is set.
    \end{itemizesquishopt}
\end{example}
%
\paragraph{Configuration of the debugger}
The debugger can be either interactive, waiting for the user to issue commands and execute them, or passive, with the program executing normally until a breakpoint or a watchpoint is triggered or the user interrupts the execution.
The debugger keeps track of the current breakpoints, watchpoints and of user's checkpoints.
\begin{definition}[Configuration of the debugger]
	\label{def:debugger}
	A \textit{configuration of the debugger} is a 4-tuple $(q_d, \Bp, \Wp,$ $ \Checkp) \in \{\Interactive, \Passive\} \times \mathcal{P}(\Breakpoints) \times \mathcal{P}(\Watchpoints)  \times \Checkpoints^*$ where
	\begin{itemizeopt}
		\itemopt[!]
		$q_d$ is the current mode of the debugger, either $\Interactive$ (interactive) or $\Passive$ (passive),
		\itemopt[!]
		$\Bp$ and $\Wp$ are the \emph{sequences of breakpoints and watchpoints} handled by the debugger,
		\itemopt[!]
		$\Checkp$ is the \emph{sequence of checkpoints} set by the user.
	\end{itemizeopt}
\end{definition}

\hideInRLYShortVersion{
Sequences are used for $\Checkp$, $\Wp$ and $\Bp$ in order to allow the user manipulate checkpoints, watchpoints and breakpoints by their index.
}

\ifdefined\HIDESCENARIOFORMALISM\else
%
\htwo{The Scenario}
\label{sec:scenario}
%
The scenario reacts to monitor events by executing actions that update the state of the program, of the debugger and of the scenario itself.
We define actions, then reactions, and finally the scenario itself.
Actions are executed when monitor events are received according to the notion of scenario reactions.

\hideInRLYShortVersion{
\begin{figure}
	\small{\[
	\begin{array}{lcll}
	  a &  ::=  &  v \; := \; e								  \; \textit{\footnotesize(assignment)} \\
		&   |   &  v \; := \; \mathtt{checkpoint}  \; \textit{\footnotesize(saving a checkpoint)} \\
		&   |   &  a \; ; \; a							 \; \textit{\footnotesize(sequential composition)} \\
		&   |   & \mathtt{if} \; e \; \mathtt{then} \; \mathrm{Action} \; \mathtt{else} \; \mathrm{Action}  \; \textit{\footnotesize(conditional statement)} \\
		&   |   & \mathtt{while} \; e \; \mathtt{then} \; \mathrm{Action} \; \mathtt{else} \; \mathrm{Action}  \; \textit{\footnotesize(loop)} \\
		&   |   & \texttt{restore-checkpoint} \; \mathrm{checkpoint} \; \textit{\footnotesize(restart a checkpoint)}  \\
		&   |   & \mathtt{setWatchpoint}   \; w \; t  \; \textit{\footnotesize(set a watchpoint. $ t \in \{\mathtt{r},\mathtt{w},\mathtt{rw}\}$)} \\
		&   |   & \mathtt{setBreakpoint}   \; b \; \textit{\footnotesize(set a breakpoint)}  \\
		&   |   & \mathtt{unsetWatchpoint} \; w \; \textit{\footnotesize(remove a watchpoint)} \\
		&   |   & \mathtt{unsetBreakpoint} \; b \; \textit{\footnotesize(remove a breakpoint)}
	\end{array}
	\]}
	with $e$ a usual expression in a programming language and $v \in \Names$.
	\caption{Grammar of Scenario actions}
	\label{fig:sagramm}
\end{figure}
}

\begin{definition}[Scenario action]
\ifdefined \RLYSHORTVERSION \else
	The set of possible actions, $\A$, is defined by the grammar in Figure~\ref{fig:sagramm}.
\fi
The set of possible actions, $\A$, is constructed like the set of statements in a classical programming language in which it is also possible to set and remove breakpoints, watchpoints and checkpoints and restore checkpoints.
\end{definition}

\begin{definition}[Scenario reaction]
	A scenario reaction is a 3-tuple $(\inmathlist{lt, q_m, a}) \in \{\entering, \leaving\} \times Q_m \times \A$, where $\mathit{lt}$ determines the ``moment of the reaction", $q_m$ is the state of the monitor to which the reaction is attached, and $a$ is an action to be executed.
	The set of scenario reactions is denoted $\ScenarioReaction$.
\end{definition}

The scenario reaction $(\inmathlist{lt, q_m, a})$ is triggered when entering (resp.) leaving state $q_m$ in the monitor when
$lt = \entering$ (resp. $lt = \leaving$.
When $(\inmathlist{lt, q_m, a})$ is triggered, action $ a $ is executed.
A scenario is specified by giving a list of reactions and an environment $m_s$ used by actions.
If a transition starting from a given state and leading to the same state is taken by the monitor, this state is both left and entered.

\begin{definition}[Scenario]
	\label{def:scenario}
	A scenario is a pair $ (m_{s}^{0}, S) \in (\Names \rightarrow \Values) \times
	\ScenarioReaction^* $ where
	$m_{s}^{0}$ is an initial environment and
	$S$ a list of scenario reactions.
\end{definition}

\begin{remark}
	$S$ is a list (and not a set) because if a state-update in the monitor triggers more than one scenario reactions, these reactions are handled in order in $S$.
\end{remark}

\ifdefined \RLYSHORTVERSION \else
\hideInRLYShortVersion{
\exref{ex:scenario} illustrates the notion of scenario.
}
\begin{example}[Scenario]
\label{ex:scenario}
Assuming that $\mathtt{a1}$ and $\mathtt{a2}$ are two scenario actions and
$\mathtt{x}$ a monitor state, the following listing describes a scenario:
\begin{lstlisting}[style=PseudoCodeStyle,escapeinside={<@}{@>}]
accesses := 0

on entering state x do
accesses := accesses + 1
if accesses = 2 then
 <@\textit{[do something]}@>
else
 <@\textit{[do something else]}@>
\end{lstlisting}

This listing defines the scenario $(\inmathlist{[\mathrm{accesses} \mapsto 0], ((\entering, \mathtt{x}, a))})$ where action $a$ increments variable $\mathtt{accesses}$ and its behavior depends on the value of variable $\mathtt{accesses}$, the environment $[\mathrm{accesses} \mapsto 0]$ is the initial memory of the scenario and $(\inmathlist{\entering, \mathtt{x}, a})$ is its only reaction.
\end{example}
\fi
\fi

\hone{Gathering the Components}
\label{sec:evolconfigurations}

\begin{figure}
	\includegraphics[width=1\linewidth]{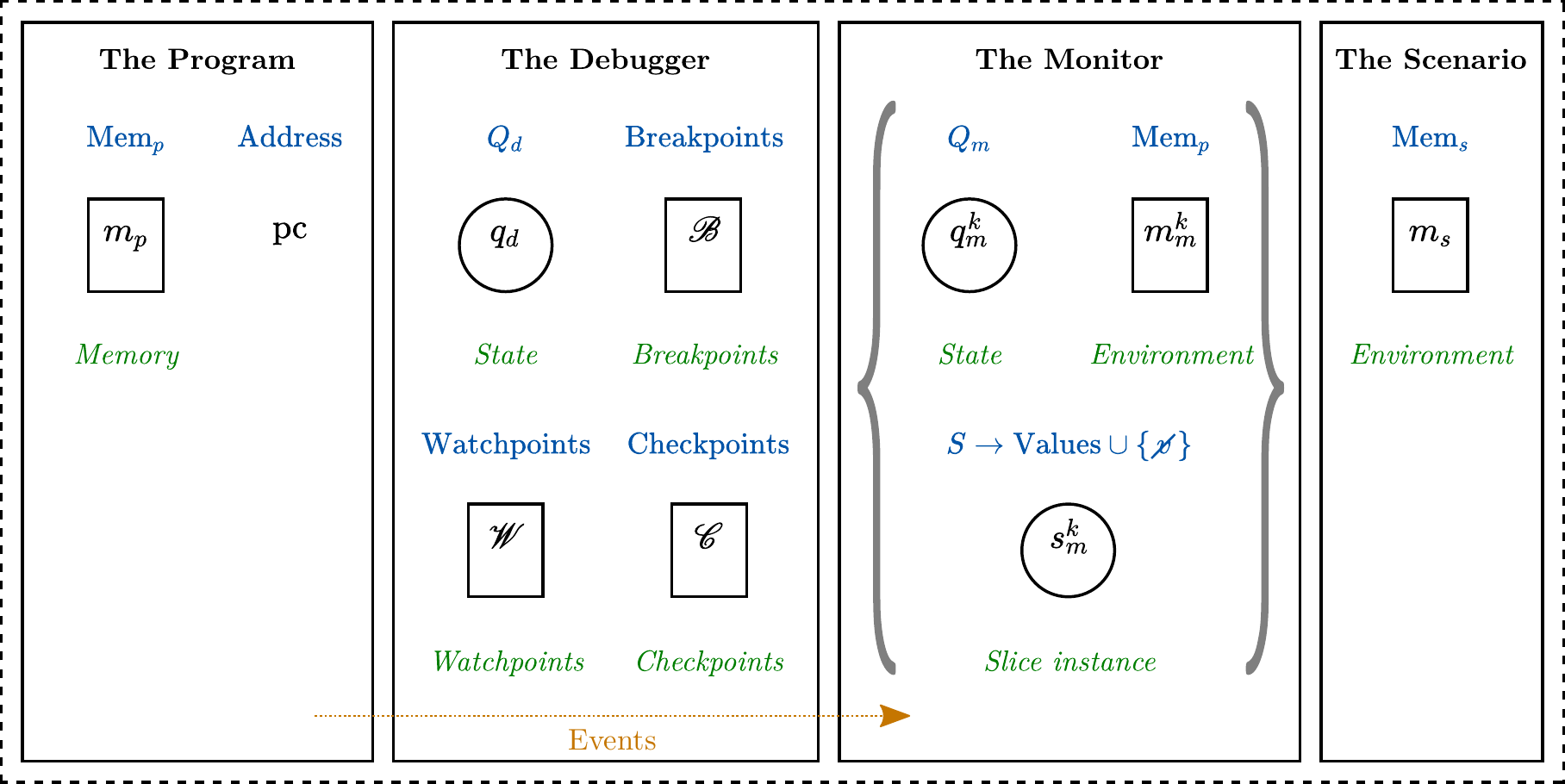}
	\vspace{-2em}
	\caption{Configuration of a \irv-program}
	\label{fig:wholesys}
\end{figure}

In this section, we give the representation of the state of the \irv-program at each execution step (i.e. its configuration). We then describe its evolution by means of pseudo-code, precisely explaining how it transitions from one configuration to another.
The \irv-program is depicted in Figure~\ref{fig:wholesys}.
Let
$P = (\inmathlist{\Sym, m_0^p , \start, \runInstr})$ be a program,
$M = (\inmathlist{Q_m , q_{m}^{0} , m_{m}^{0} , \Sigma_m , \Delta_m}) $ a monitor and
$ S = (m_s^0, S)$ a scenario.
The {\irv}-program, denoted by $\irv(P,M,S)$, is defined as the composition of $P$, $M$ and $S$ synchronized on events.
We first define the configurations of the {\irv}-program in \secref{sec:concepts:configuration} and the evolution of its configurations in \secref{sec:concepts:evolution} driven by the instrumentation functions of debugger defined in \secref{sec:instr}.
%
\htwo {Configuration of the Composition}
\label{sec:concepts:configuration}
%
We define the configurations of the {\irv}-program.
\begin{definition}[Configuration of the {\irv}-program]
    \label{def:confsys}
	 A \textit{configuration of $\irv(P, M, S)$} is a 4-tuple $\wholeState = (\inmathlist{c_{\mathrm{P}}, c_{\mathrm{dbg}}, c_{\mathrm{M}}, c_{\mathrm{S}}}) \in (\Mem \times \Address) \times (\{\Interactive, \Passive\} \times \Breakpoints^* \times \Watchpoints^* \times \Checkpoints^*) \times C_m \times \Mem_s$.
	The initial configuration of $\irv(P, M, S)$ is
$ \wholeState^0 = (\inmathlist{(m_{p}^{0}, \pc_0), (\Interactive, \varepsilon, \varepsilon, \varepsilon), \{(\init, m_{m}^{0}, * \mapsto \uninstantiated, \varnothing)\}, m_{s}^{0}}) $.
\end{definition}
A configuration is composed of the initial program memory,
the start address of the program as the PC,
the debugger is interactive and does not manage any breakpoint, watchpoint or checkpoint,
the monitor has one slice instance of the property that is in its initial state and in
its initial environment and all parameters of the slice are uninstantiated
and the memory of scenario is its initial memory.

\htwo{Instrumentation Functions of the Debugger}
\label{sec:instr}

\hideInRLYShortVersion{
\myparag{$\setBP$, $\unsetBP$}
}
\hideInRLYShortVersion{
Breakpoints and watchpoints are used to monitor function calls.
We define the following functions:~$\setBP$ (sets a breakpoint),
$\unsetBP$~(find a breakpoint by its address and remove it from the memories of
the program and the debugger), $\setWP$~(sets a watchpoint) and
$\unsetWP$~(unsets a watchpoint).

To set a breakpoint, we need to write a special instruction in the program memory.
\hideInRLYShortVersion{
When this instruction is encountered during the execution, the execution is suspended and the debugger takes control over it.
}
We also need to keep the word we replace in memory, so when the execution is resumed from this breakpoint or when the breakpoint is removed, the special breakpoint instruction is replaced by the stored instruction in the program memory.
}

\hideInRLYShortVersion{
Several breakpoints can be set at the same address.
For instance, the monitor and the user might want to set a breakpoint on the
same function.
Breakpoints have to be stored in order in the structures of the debugger.
We therefore use a list to save them.
}

We define $ \setBP $, a function that sets a breakpoint and register it in the configuration of the debugger.
We indicate if the breakpoint was set by the user or by the monitor,
so that breakpoints set by the user do not call the monitor and breakpoint set
by the monitor are not seen by the user.
We define $ \setInstrBp : \Env \times \Address \rightarrow \Env $, a function
replacing the word at a given address in the program memory by
a breakpoint instruction (denoted by $\BREAK$).
$ \setInstrBp(m_p, \addr)[\addr] = \BREAK $ and $\forall a \in \Address : a \neq \addr,
\setInstrBp(\inmathlist{m_p, a})[a] = m_k[a] $.
Function $\setBP : \Mem \times \Breakpoints \times \Address \times
\B $ sets a breakpoint and saves it in the memory of the debugger:
$ \setBP(\inmathlist{m_p, \Bp, \addr, \isuserbp}) = (\inmathlist{\Bp', m_p'}) $ with
$ m_p' = \setInstrBp(\inmathlist{m_p, \addr}) $ and
$ \Bp' = (\inmathlist{\addr, m_p[\addr], \isuserbp}) \cdot \Bp $.
In the same way, we define $ \setWP:
\Watchpoints \times \Address \times \B \times \B \times \B$ that adds
a watchpoint in the memory of the debugger:
$ \setWP(\Wp, \addr, r, w, \isUserWatchpoint) = (\addr, r, w,
\isUserWatchpoint) \cdot \Wp $.
For watchpoints, the program memory does not need to be modified.\hideInRLYShortVersion{
\myparag{unsetBP, unsetWP}
}
We define $ \unsetBP $, which unsets a breakpoint and stops keeping track of it.
$ \unsetBP(\inmathlist{m_p, \Bp \setminus \{bp\}, a, \isuserbp}) = (\inmathlist{m_p', \Bp', b}) $ s.t. $bp$ is a breakpoint in $\Bp$ matching $ (\addr, \_, \isuserbp) $ and
\ifdefined \SHORTVERSION
$\forall \addr \in \Address, m_p'[\addr] = \instr $ s.t. $(\_, \instr, \_) = bp$ or $m_p'[\addr] = m_p[a] $ otherwise.
\else
\[
	\forall \addr \in \Address, m_p'[\addr] = \begin{cases}
	\instr \, & \text{s.t.} \, (\_, \instr, \_) = bp \\
	m_p[a] & \text{otherwise}
	\end{cases}
\]
\fi
We also define $\unsetWP(\inmathlist{\Wp, a, r, w, \isuserwp}) = (\inmathlist{\Wp \setminus \{wp\}, wp}) $ s.t. $wp$ is a watchpoint in $\Wp$ matching $(\inmathlist{\addr, r, w, \isuserwp})$.
\hideInRLYShortVersion{
\begin{remark}
Functions $\setInstrBp$, $\setBP$, $\unsetBP$ model the behavior of a
real debugger using software breakpoints. A software breakpoint is implemented
using a trap instruction. When several breakpoints are set at a single address,
the debugger sets one trap instruction at this address in the memory of the
program and keeps track of all the breakpoints in its internal structures.
\end{remark}
}
For each monitor update, breakpoints and watchpoints corresponding to events monitored by the old (resp. new) state must be unset (resp. set).
\hideInRLYShortVersion{
\myparag{$\instrument$, $\unInstrument$}
}
In \algoref{algo:setbp}, we define function $ \instrument $ used to
set breakpoints needed for the new state.
It takes the old program memory, the new state and the symbol table and returns a new memory and a new list of breakpoints.
Function $ \unInstrument $ unsets breakpoints that are not needed anymore.

\begin{remark}
Instrumenting to monitor value changes for an expression may require setting several watchpoints. The list of watchpoint to set for an expression is returned by function $\mathrm{variablesAccesses} : \Events \rightarrow \mathcal{P}(\Watchpoints) $ which is not defined formally here for the sake of simplicity.
\end{remark}

\begin{algorithm}[!h]
	\caption{$\instrument$, $\unInstrument$}
	\label{algo:setbp}
	\begin{algorithmic}[1]
        \ifdefined\RLYSHORTVERSION
            \renewcommand{\needsBreakpoint}[1]{\type(#1) = \FunctionCall}
        \else
		\Function{needsBreakpoint}{e} \Commentopt{Instrumenting function calls uses breakpoints}
			\State \Return $ \type(e) = \FunctionCall $
		\EndFunction
        \fi
		\Function{needsWatchpoint}{e} \Commentopt{Instrumentation of value accesses needs watchpoints}
			\State \Return $ \type(e) \in \{\ValueWrite\, \ValueRead, \UpdateExpr\} $
		\EndFunction

		\Function{\wpEvtToAddressesnorm}{e, Sym}
			\State $ res \gets \emptyset $ \Commentopt{Set of watchpoints needed to monitor the expression given by the event}
			\ForAll{$(n, r, w) \in \mathrm{variablesAccesses}(e)$} \Commentopt{We assume the existence of $\mathrm{variablesAccesses}$}
				\State $ res \gets res \cup \{(\Sym[n], r, w)\} $
			\EndFor
			\State \Return $res$
		\EndFunction

		\Function{\unInstrumentnorm}{$m_p, \Bp, \Wp, M, Sym$}
			\State \Let $m_p' \gets m_p$ \Commentopt{To remove instrumentation for the current state of the monitor}
			\ForAll{$e \in \enabled(M)$} \Commentopt{For each possible event}
				\If{$\needsBreakpoint{e}$} \Commentopt{We remove the corresponding breakpoint, if any}
					\State $\Bp' = \Bp \setminus \{b \; | \; b \; \textrm{matches} \; (\addr, \_, \isuserbp) \}$
				\EndIf

				\If{$\needsWatchpoint(e)$} \Commentopt{We remove the corresponding watchpoints, if any}
					\ForAll{$(a, r, w) \in \wpEvtToAddresses(e, \Sym)$}
						\State $\Wp' \gets
						\unsetWP(\Wp', a, r, w, \false)$
					\EndFor
				\EndIf
			\EndFor
			\State \Return{$(m_p', \Bp', \Wp')$}
		\EndFunction

		\Function{\instrumentnorm}{$m_p, \Bp, \Wp, M, Sym$}
			\State \Let $(m_p', \Bp', \Wp') \gets (m_p, \Bp, \Wp)$ \Commentopt{To instrument for the current state of the monitor}
			\ForAll{$e \in \enabled(M)$} \Commentopt{For each possible event}
				\If{$\needsBreakpoint{e}$} \Commentopt{We create the eventually needed breakpoint}
					\State $(m_p', \Bp') \gets \setBP(m_p', \Bp', \bpEvtToAddress(e, \Sym), \false)\!\!$
					\label{algoline:instrumentcalltosetbp}
				\EndIf
				\If{$\needsWatchpoint(e)$} \Commentopt{We create the eventually needed watchpoints}
					\ForAll{$(a, r, w) \in \wpEvtToAddresses(e, \Sym)$}
						\State $\Wp' \gets
						\textsc{setWP}(\Wp', a, r, w, \false)$
					\EndFor
				\EndIf
			\EndFor
			\State \Return{$(m_p', \Bp', \Wp')$}
		\EndFunction
	\end{algorithmic}
\end{algorithm}
\hideInRLYShortVersion{
\myparag{removeAllBPs}
}

In a checkpoint, the saved program memory must not contain any instruction $\BREAK$.
A function to remove all breakpoints from the memory is therefore needed.
We define $ \removeAllBPs $, a function which iterates over the list of breakpoints of the debugger and replaces each instruction $\BREAK$ by the original instruction. $\removeAllBPs(m_p, \epsilon) = m_p$ and
$ \inmathlist{\forall \Bp  \in \Breakpoints: \Bp \neq \epsilon, \removeAllBPs(m_p, \Bp) = \removeAllBPs(m_p', \Bp')} $ where
$ \Bp = (\addr, \instr, b) \cdot \Bp'$ and $ m_p' = m_p[\addr \mapsto \instr]$.
When a checkpoint is restored, current breakpoints must be set in the memory.
We therefore define the function $\restoreAllBPs$ which iterates over the list of breakpoints and sets the instruction $\BREAK$ at the relevant addresses.
$\restoreAllBPs(\inmathlist{m_p, \Bp}) = m_p'$ with
\ifdefined\SHORTVERSION
	$\forall a \in \Address, m_p'[a] = \BREAK $ if $\exists (\inmathlist{\addr, \instr, b}) \in \Bp : \addr = a$, $ m_p[a] $ otherwise.
\else
\[
	\forall a \in \Address, m_p'[a] = \begin{cases}
	\BREAK & \text{if } \, \exists (\addr, \instr, b) \in \Bp : \addr = a \\
	m_p[a] & \text{otherwise}
	\end{cases}
\]
\fi
%
\htwo{Evolution of the \irv-program}
\label{sec:concepts:evolution}
%
{
	\def\OldComma{,}
	\catcode`\,=13
	\def,{%
		\ifmmode%
		\OldComma\discretionary{}{}{}%
		\else%
		\OldComma%
		\fi%
	}%
In this section, we describe the precise behavior of the \irv-program.
The algorithm describing the general behavior of the \irv-program is given in \algoref{algo:general} and explained right after.
The initial configuration of the \irv-program is $ ((m_{p}, \pc), (q_d, \Bp, \Wp, \Checkp),
M\commams) = ((m_{p}^{0}, \pc_0), (\Interactive, \varepsilon, \varepsilon, \varepsilon),
\{(\init, m_{m}^{0}, * \mapsto \uninstantiated, \varnothing)\}, m_{s}^{0}) $.
In this configuration, the debugger is in interactive mode, meaning it is waiting
for commands from the user (Line~\ref{algline:init})
}

\begin{algorithm}[t]
	\caption{Behavior of the System}
	\label{algo:general}
	\begin{algorithmic}[1]
		\State \Let $ cont \gets \true $
		\While{$cont$}
			\If{$q_d = \Interactive$} \label{algline:init} \Commentopt{Interactive execution}
				\State $(cont, ((m_p, \pc), \wholeState) \gets \textsc{handleUserCMD}(\wholeState)$
			\ElsIf{User stops the execution or $m_p[\pc] = \stopInstr$} \label{algoline:interrupt} \Commentopt{Interruption}
				\State $q_d \gets \Interactive$
			\ElsIf{$m_p[\pc] \neq \BREAK \land  m_p[\pc] \neq \stopInstr$} \label{algoline:normalexec} \Commentopt{Normal execution}
				\State $\wholeState \gets \normalStep(\wholeState)$
			\ElsIf{$m_p[\pc] = \BREAK$} \label{algoline:dbginstr}  \Commentopt{A breakpoint is reached}
				\State $ \wholeState \gets \handleBP(\wholeState) $
			\EndIf
		\EndWhile
	\end{algorithmic}
\end{algorithm}

\begin{algorithm}[!t]
	\caption{Behavior when the debugger is Interactive}
	\label{algo:interactive}
	\begin{algorithmic}[1]
		\Function{handleUserCMD}{$\wholeState$}
			\State \Let $cont \gets \true$
			\Switch{getUserCMD()}
				\Case{$\mathtt{load \; monitor}$}  \label{algoline:loadmon} \Commentopt{Loading the monitor}
					\State $(m_p, \Bp, \Wp) \gets \instrument(m_p, \Bp, \Wp, M, \Sym)$
				\EndCase
				\Case{$\mathtt{restart}\, n$} \label{algoline:restart} \Commentopt{Restarting from a checkpoint}
					\State $ (m_p', \Bp, \Wp) \gets \unInstrument(m_{p}^{tmp}, \Bp, \Wp, M, \Sym) $
					\State $ (m_{p}^{tmp}, pc, M) \gets \Checkp_n $
					\State $ (m_p, \Bp, \Wp) \gets \instrument(\restoreAllBPs(m_{p}^{tmp}, \Bp),$
					\State \hide{$(m_p, \Bp, \Wp) \gets \instrument($}$\Bp, \Wp, M, \Sym) $
				\EndCase
				\Case{$\mathtt{continue}$} \label{algoline:continue} \Commentopt{continuing the execution}
					\State $\wholeState \gets \textsc{interactiveStep}(\wholeState)$
					\ifdefined \SHORTVERSION ; \else \State \fi $q_d \gets \Passive$
				\EndCase
				\Case{$\mathtt{break} \; a$} \label{algoline:setbreak} \Commentopt{Setting a breakpoint}
                    \TernaryAss{$ (m_p, \Bp) $}{$a \in \Address$}{$\setBP(m_p, a, \Bp, \true)$}{$\setBP(m_p, \Sym(name), \Bp, \true)$}
				\EndCase
				\Case{$\mathtt{watch} \, mode \; a, a \in \Address$} \label{algoline:setwatch} \Commentopt{Setting a watchpoint}
					\State 	$\Wp \gets \Wp \cdot (a, \mathtt{r} \in mode , \mathtt{w} \in mode, \true)$
				\EndCase
				\Case{$\mathtt{checkpoint}$} \label{algoline:setcheck}  \Commentopt{Setting a checkpoint}
					\State $(n, {\Checkp}_N) \gets (\min N, \bot) $\
					\ifdefined \SHORTVERSION
						\State $\Checkp_n \gets (\removeAllBPs(m_p, \Bp), M, \pc) $
					\else 
						\State $\forall k \in \mathbb{N}, \Checkp_k \gets \begin{cases}
							\Checkp_k & k \neq n \\
							(\removeAllBPs(m_p, \Bp), M, \pc) &  k = n
						\end{cases}$
					\fi
				\EndCase
				\ifdefined\RLYSHORTVERSION
				\Case{$\mathtt{step}$:$\;\wholeState \gets \textsc{interactiveStep}(\wholeState)$} \label{algoline:setstep}
				\EndCase
				\Case{$\mathtt{exit}$:$\;cont \gets \false$}
				\EndCase
				\else
				\Case{$\mathtt{step}$} \label{algoline:setstep}  \Commentopt{Executing a step}
					\State $\wholeState \gets \textsc{interactiveStep}(\wholeState)$
				\EndCase
				\Case{$\mathtt{exit}$} \Commentopt{Stopping the \irv-program}
					\State $cont \gets \false$
				\EndCase
				\fi
				\hideInShortVersion{
				\Case{other}
					\State print ``Illegal Command''
				\EndCase
				}
			\EndSwitch
			\State \Return $(cont, \wholeState)$
		\EndFunction
	\end{algorithmic}
\end{algorithm}

\myparag{First step of the execution}
When starting the execution of the \irv-program, the monitor is initialized (command $\mathrm{load \; monitor}$, \algoref{algo:interactive}, Line~\ref{algoline:loadmon}): breakpoints and watchpoints relevant to the initial state of the property are set.

\myparag{Normal execution}
If the debugger is passive and the instruction about to be executed is not
an instruction $\BREAK$),
the program executes normally as if there were no debugger and no monitor (\algoref{algo:general}), Line~\ref{algoline:normalexec}.
In function $\normalStep$ (\algoref{algo:step}, Line~\ref{algoline:normalstep}),
the PC and the program memory are updated by function $\runInstr$ which runs the instruction to be executed.
Watchpoints relevant to memory accesses made by this execution are handled.
The instruction $\stopInstr$ ends the execution~(\algoref{algo:general}, Line~\ref{algoline:interrupt}).

\myparag{Handling a watchpoint}
In \algoref{algo:handlewp}, we define $\handleWP$. If the watchpoint was
set by the user, the state of the \irv-program is returned as is, except for the state
of the debugger, which becomes interactive.
If the watchpoint belongs to the monitor, the corresponding events are applied using the function $\applyEvents$ defined
in \algoref{algo:applyEvent}, updating the monitor\ifdefined\HIDESCENARIOFORMALISM\else and executing the scenario\fi.

\myparag{The user sets a breakpoint}
When the debugger is interactive ($\Interactive$), the user can set a breakpoint~(\algoref{algo:interactive}, Line~\ref{algoline:setbreak}) by giving either an address in the program memory or a symbol (function) name transformed into an address using the symbol table $\Sym$, part of the definition of the program.
If the user issues a command to set a breakpoint at address $a$, the function $\setBP$ updates the current program memory $m_p$ and the list of breakpoints $\Bp$ of the debugger.
The resulting memory $m_p'$ and list of breakpoints $\Bp'$ are stored in the configuration of the \irv-program.

\myparag{The user sets a watchpoint}
In interactive mode ($\Interactive$), the user can set
a watchpoint by giving the address in the program memory where it should be
set~(\algoref{algo:interactive}, Line~\ref{algoline:setwatch}).

\myparag{The user sets a checkpoint}
In interactive mode, the user can set a checkpoint~(\algoref{algo:interactive}, Line~\ref{algoline:setcheck}).
Several objects are saved: the program memory (without the breakpoints instructions), the PC and the state of the monitor.
The least identifier $n$ that is not used for a checkpoint is used as an index for the new checkpoint. The checkpoint is stored in $\Checkp'_n$.

\myparag{The user restarts a checkpoint}
In interactive mode ($\Interactive$), the user can restore
a checkpoint~(\algoref{algo:interactive}, Line~\ref{algoline:restart}).
The current program memory, the current PC, the current configuration of the monitor and its current memory are restored from the checkpoint.
Current breakpoints are set in the newly restored program memory (this behavior matches the behavior of GDB and LLDB).

\myparag{A breakpoint instruction is encountered}
When encountering a breakpoint instruction, the debugger has to check if it matches a breakpoint of the user or a breakpoint of the monitor.
In the first case, the \irv-program transitions to the $\Interactive$ state.
In the second case, the event is applied.

\begin{remark}
In real systems, the breakpoint instruction triggers a trap caught by the operating system
which suspends the execution and informs the debugger of the trap.
Traps are not described in our model because we do not model the OS.
The behavior of our model is otherwise close to the reality.
\end{remark}

\begin{algorithm}[!t]
	\caption{Handling Instrumentation (generating events)}
	\label{algo:handlebp}
	\label{algo:handlewp}
	\begin{algorithmic}[1]
		\newcommand{\access}{\mathrm{access}}
		\newcommand{\watchpointsMatchingNorm}{watchpointsMatching}
		\newcommand{\watchpointsMatching}{\mathrm{\watchpointsMatchingNorm}}

		\Function{\handleBPnorm}{$\wholeState$}
		\If {$\exists \, \instr : (\addr, \instr, \true) \in \Bp$}
			\Commentopt{There is a breakpoint of the user}
			\State \Return $(m_p, \pc), (\Interactive, \Bp, \Wp, \Checkp), M\commams)$
		\EndIf
        \Commentopt{The breakpoint is a breakpoint of the monitor}
        \State \Return $ \applyEvents( \bpToEvts(m_p', \pc, M, Sym), \wholeState ) $ \Commentopt{We update the monitor\ifdefined\HIDESCENARIOFORMALISM \else and run the scenario\fi}
		\EndFunction

		\Function{\watchpointsMatchingNorm}{$\Wp, (\addr, r, w)$} \Commentopt{Does an access match a watchpoint?}
			\State $Wp_s \gets \emptyset$
			\ForAll{$(\addr', r', w', \isUserWatchpoint) \in \Wp$} \Commentopt{For each watchpoint}
				\If{$\addr = \addr' \land (r = r' \lor w = w')$} \Commentopt{If the access matches, we retain it}
					\State $Wp_s \gets \Wp_s \cup \{(\addr', r', w', \isUserWatchpoint)\}$
				\EndIf
			\EndFor
			\State \Return $Wp_s$ \Commentopt{We return all retained watchpoint}
		\EndFunction
		\Function{\handleWPnorm}{$\accesses, \wholeState$}
			\State $\Wp_s \gets \emptyset$
			\ForAll{$\access \in \accesses$}
				\State $\Wp_s \gets  \Wp_s \cup \{\watchpointsMatching(\Wp, \access)\}$
			\EndFor
			\If {$\exists (\_, \_, \_, \true) \in \Wp_s$}
				\Commentopt{There is a watchpoint of the user}
				\State \Return $((\pc, m_p), (\Interactive, \Bp, \Wp, \Checkp), M\commams)$
			\EndIf
			\Commentopt{The watchpoint is a watchpoint of the monitor}
			\State \Return $ \applyEvents(\wpsToEvts(Wp_s, P, M, Sym),\wholeState) $ \Commentopt{We update the monitor\ifdefined\HIDESCENARIOFORMALISM \else and run the scenario\fi}
		\EndFunction
	\end{algorithmic}
\end{algorithm}

\myparag{Handling a breakpoint}
In \algoref{algo:handlebp}, we define $\handleBP$. If the breakpoint belongs to the user, the \irv-program becomes interactive but is not otherwise modified.
If the breakpoint belongs to the monitor, breakpoints are removed from the program memory,
a corresponding before event is applied using the function $\applyEvent$ defined in \algoref{algo:applyEvent},
the original instruction is run and an after event is applied using the function $\applyEvent$ that updates the state of the \irv-program.
It first updates each slice of the configuration of the monitor according to the event and the transition relation, retrieving the set of transitions involved.
\ifdefined\HIDESCENARIOFORMALISM \else
It then applies the scenario using the function $\applyScenario$~(applyScenario) defined in \algoref{algo:applyScenario}.
The scenario can update the whole state of the \irv-program. It is applied only if the current state has been
updated~(Line~\ref{algoline:aeas}) of \algoref{algo:applyEvent}).
For each entry of the scenario, if the event corresponds to the entry, the corresponding action is run with function runAction.
For the sake of conciseness, function runAction is not defined precisely here. See \algoref{algo:general}, Line~\ref{algoline:dbginstr}.
\fi
Breakpoints are restored and the instrumentation needed for the new current state is added.

\begin{algorithm}[t]
	\caption{Handling events}
	\label{algo:applyEvent}
	\begin{algorithmic}[1]
        \Function{\updatemonnorm}{$M, e$}:
            \State $ M' \gets \emptyset $
            \ifdefined\HIDESCENARIOFORMALISM \else
                \ifdefined\RLYSHORTVERSION
                    ;
                \else
                    \State
                \fi $ trans \gets \emptyset $
            \fi
            \ForAll{$ (q, m, s, sp) \in M $}
                \ForAll{$ (sb, q_s, e_t, g, upd, q_d) \in \Delta_m $}
                    \State $instance \gets [s \mapsto (\values(e))(p) \; | \; \exists \, (p, s) \in sb]$
                    \If{$ q = q_s \land s \sqsubseteq instance \land \nexists (q', m', s', sp') \in M : s \sqsubset s' \land s' \sqsubseteq instance $}
                        \ifdefined\HIDESCENARIOFORMALISM
                            \TernaryAss{$M'$}{$s \sqsubset instance$}{$M' U (q_d, upd(\values(e), m), instance, s)$}{$M' U (q_d, upd(\values(e), m), s, sp)$}
                        \else
                            \If{$s \sqsubset instance$} 
                                \State $ M' \gets M' U (q_d, upd(\values(e), m), instance, s) $
                                \State $ trans \gets trans \cup (q_s, q_d) $
                            \Else
                                \State $ M' \gets M' U (q_d, upd(\values(e), m), s, sp) $
                            \EndIf
                            \State $ trans \gets trans \cup (q_s, q_d) $
                        \fi
                    \Else
                        \State $ M' \gets M' \cup (q, m, s, sp) $
                    \EndIf
                \EndFor
            \EndFor
            \ifdefined\HIDESCENARIOFORMALISM
                \State \Return $M'$
            \else
                \State \Return $(M', trans)$
            \fi
        \EndFunction
		\Function{applyEvent}{$\wholeState, e$} \Commentopt{We apply one before or after event}
            \ifdefined\HIDESCENARIOFORMALISM
                \State \Return $\updateMon(M, e)$
            \else
                \State $(M', \mathit{trans}) \gets \updateMon(M, e)$
                \State \Return $\applyScenario(S, \wholeState, M', \mathit{trans})$
                \label{algoline:aeas} \Commentopt{We apply the scenario}
            \fi
		\EndFunction
		\Function{\applyEventsnorm}{$evList, \wholeState$} \Commentopt{For a list of generated events}
			\State $ m'_p \gets \removeAllBPs(m_p, \Bp) $ \label{algoline:aermallbp} \Commentopt{Step 1: remove all instrumentation}
			\State $ (\Bp', Wp') \gets \unInstrument(\Bp, \Wp, M, \Sym) $~\label{algoline:aermmonbp}
			\State $(\pc', q_d', \Checkp', M'\commamsprime) \gets (\pc, q_d, \Checkp, M\commams)$
			\ForAll{$ e \in evList $ s.t. $ \isBefore(e) $} \label{algoline:aebefore}  \Commentopt{Step 2: apply before events}
				\State $ \wholeState' \gets \applyEvent(\wholeState', e) $
			\EndFor
			\State $ (m_p', \pc')  \gets \runInstr(m_p', \pc') $ \label{algoline:aeri} \Commentopt{Step 3: run the instruction}
			\ForAll{$ e \in evList $ s.t. not $ \isBefore(e) $}  \Commentopt{Step 4: apply after events}
				\State $ \wholeState \gets \applyEvent(\wholeState\commamsprime, e) $
			\EndFor
			\State $ {m'}_p^{tmp}  \gets \restoreAllBPs(m_p', \Bp') $ \label{algoline:aeafter} \Commentopt{Step 5: restore / update instrumentation}
			\State $ (m_p', \Bp', \Wp')  \gets \instrument({m'}_p^{tmp}, \Bp', \Wp', M', Sym) $ \label{algoline:aeinstr}
			\State \Return $\wholeState'$
		\EndFunction
	\end{algorithmic}
\end{algorithm}

\ifdefined\HIDESCENARIOFORMALISM \else
\begin{algorithm}[!t]
	\caption{$\textsc{applyScenario}$}
	\label{algo:applyScenario}
	\begin{algorithmic}[1]
		\Function{\scenarioReactionMatchesEvent}{$lt, q_m, q_s, q_m'$}
			\State \Return $(lt = \leaving \; \land \; q_m = q_s) \lor (lt = \entering \; \land \; q_m' = q_s)$
		\EndFunction
		\Function{$\applyScenario$}{$s, P, D, M^i\commams, M^f$}
			\State $((m_p', \pc'), D'\commamsprime) \gets (P, D\commams) $

			\ShortIf{$s = \epsilon$}{\Return $(m_p', \pc'), D'\commamsprime$} \Commentopt{End of scenario reached (or empty scenario)}

			\ForAll{$(q_m', m_m', s', sp') \in M'$} \Commentopt{For each slice in the monitor}
				\State \Let $(q_m, m_m, s) $ \textbf{be such that} \Commentopt{We get the previous state of the slice}
				\State $ \quad s = s' \land (q_m, m_m, s', sp') \in M $ \textbf{or else}
				\State $ \quad s = sp' \land \exists \, sp \, : \, (q_m, m_m, sp', sp) \in M $ \textbf{or else}
				\State $ \quad (q_m, m_m, s) = (\varnothing, \varnothing, \varnothing)$ \Commentopt{The previous state may not exist}
				\If{$q_m \neq q_m'$} \Commentopt{If the state of the slice changed}
					\State $(lt, q_s, a) \gets \mathrm{head(s)}$ \Commentopt{We apply the current scenario reaction}
					\If{\textsc{\scenarioReactionMatchesEvent}($lt, q_m, q_s, q_m'$)}
						\State ($P', D'\commamsprime) \gets \mathrm{runAction}(P, D, a)$ \Commentopt{If relevant to the current states}
					\EndIf
				\EndIf
			\EndFor
			\State ($P', D'\commamsprime) \gets \applyScenario(\mathrm{tail}(s), P', D', M^i\commams, M^f) $ \Commentopt{Recursive call}
		\EndFunction
	\end{algorithmic}
\end{algorithm}
\fi

\hideInShortVersion{
Function $ \bpToEvts $ generates a list of events from a breakpoint.
The type, the name, the parameters and the values of each event from the list is determined by the current PC, the symbol table and the program memory and is given by the instruction set of the program.
If the PC is at the first (resp. last) instruction of a function, a before (resp after) event of type $\FunctionCall$ is generated.
The function $ \bpToEvts $ specifies for each generated event whether the event
is a before event or an after event.
}

\myparag{The execution of the program is done step by step}
In interactive mode, function $\mathrm{interactiveStep}$ (\algoref{algo:step}) is executed when the command
$\mathtt{step}$ is issued. The instruction at the current address is run normally and possible watchpoints are handled~(Lines~\ref{algoline:stepbystepbreak} and Line~\ref{algoline:step}).
If the instruction is a breakpoint, the breakpoint is handled if it is set by the monitor or ignored otherwise, and the original instruction is executed.

\begin{algorithm}[t]
	\newcommand{\access}{\mathrm{access}}
	\newcommand{\watchpointsMatchingNorm}{watchpointsMatching}
	\newcommand{\watchpointsMatching}{\mathrm{\watchpointsMatchingNorm}}
	\caption{Handling a step}
	\label{algo:step}
	\label{algo:handlestepbp}
	\begin{algorithmic}[1]
		\Function{handleStepWP}{$\accesses, (m_p^{\mathrm{tmp}}, \pc^{\mathrm{tmp}}), \wholeState)$}
			\State $\Wp_s \gets \emptyset$ \Commentopt{Watchpoint during an interactive step}
			\ForAll{$\access \in \accesses$} \Commentopt{We retain each known watchpoint corresponding to}
				\State $\Wp_s \gets  \Wp_s \cup \{\watchpointsMatching(\Wp, \access)\}$ \Commentopt{an access returned by $\runInstr$}
			\EndFor
			\State $\Wp_s \gets \{(\addr, r, w, \isUserWatchpoint) \in \Wp_s \, | \, \isUserWatchpoint = \false\}$
			\Commentopt{We ignore user's watchpoints}
			\ShortIf{$\Wp_s = \emptyset$}{\Return $(m_p^{\mathrm{tmp}}, \pc^{\mathrm{tmp}}), (\Passive, \Bp, \Checkp), M\commams)$}

			\State \Return $ \applyEvents(\wpsToEvts(\Wp_s, m_p, \pc, M, Sym), \wholeState) $
			\Commentopt{We update the monitor\ifdefined\HIDESCENARIOFORMALISM\else and run the scenario\fi}
		\EndFunction
		\Function{\handleStepBPnorm}{$ ((m_p, \pc), D, M\commams)$} \Commentopt{Breakpoint during an interactive step}
			\If{$\exists \instr : (\pc, \instr, \false) \in \Bp$} \label{algoline:handlestepmonbp} \Commentopt{If a breakpoint belongs to the monitor}
				\State \Return $ \applyEvents(\bpToEvts(m_p', \pc, M, Sym), \wholeState) $
				\Commentopt{We update the monitor\ifdefined\HIDESCENARIOFORMALISM\else and run the scenario\fi}
			\EndIf \label{algoline:handlestepuserbp} \Commentopt{Otherwise, the breakpoint belongs to the user}
            \State \Let $ \instr : (\pc, \instr, \_) \in \Bp$ such that $\instr \neq \BREAK$ \Commentopt{We remove it from the memory}
            \State $\wholeState' \gets \textsc{normalStep}(\wholeState)$ \Commentopt{Execution step}
            \If{$\exists \instr : (\pc, \instr, \true) \in \Bp'$} \Commentopt{We restore the breakpoint if still present}
                \State \Return $((m'_p[\pc \mapsto \BREAK], \pc'), D', M'\commamsprime)$
            \EndIf
            \State \Return $((m'_p, \pc'), (q_d', \Bp', \Wp'), M'\commamsprime)$
		\EndFunction
		\Function{\normalStepnorm}{$\wholeState$} \label{algoline:normalstep} \Commentopt{One step of normal execution}
			\State $ (m_p^{\mathrm{tmp}}, \pc^{\mathrm{tmp}}, \accesses) \gets \runInstr(m_p, \pc) $ \label{algoline:normalstepruninstr}  \Commentopt{The current instruction is run}
			\State \Return $ \handleWP(\accesses, \wholeState) $
			\label{algoline:normalstephandlewp}
			\Commentopt{A watchpoint may be reached}
		\EndFunction
		\Function{\interactiveStepnorm}{$\wholeState$} \Commentopt{We make an interactive step}
            \ShortIf{\label{algoline:stepbystepbreak} $m_p[\pc] = \BREAK$}{\Return $ \handleStepBP(\wholeState) $ \Commentopt{We handle it}}
            \Commentopt{If a breakpoint is present}
			\If{$m_p[\pc] \neq \stopInstr$} \Commentopt{Otherwise, we execute one step normally}
                \label{algoline:step}
                \State \Return $\normalStep((m_p, \pc), (\Passive, \Bp, \Checkp), M\commams)$
            \EndIf
			\State print ``Illegal Command''
			\Commentopt{Making a step when at the end of the program is forbidden}
		\EndFunction
	\end{algorithmic}
\end{algorithm}

\myparag{The execution of the program is interrupted by the user}
The debugger switches from passive ($\Passive$)
to interactive ($\Interactive$) mode~(Line~\ref{algoline:interrupt}).
\hideInShortVersion{
\begin{remark}
}
This mechanism is meaningful if a step in the algorithm is assumed to take a non-zero amount of time.
\hideInShortVersion{
\end{remark}
}

\myparag{The execution of the program is resumed}
\hideInRLYShortVersion{
This mechanism is the inverse of the previous one.
}
If the execution is continued~(e.g. by issuing the command
$\mathtt{continue}$, see Line~\ref{algoline:continue}), a step is executed (in case the execution was interrupted by a breakpoint or a watchpoint) and the \irv-program transitions from
$\Interactive$ to $\Passive$.

\ifdefined \SHORTVERSION \else
\hzero{Preserving the Initial Program Executions}

In this section, intuitively we show that instrumenting the program for interactive runtime verification does not interfere with the program.
That is, we show that there is some relation between and the execution of the program instrumented for interactive runtime verification and the initial execution of the program.
Such a relation serves the purpose of ensuring that one observes and studies bugs that are present only in an execution of the original program.
This ensures soundness when finding bugs and reporting verdicts with a monitor.

In the following, we formalize and prove this relation.

\newcommand{\runPrgm}{\mathrm{run}(P)}

We consider the execution of a program $P$~(Definition~\ref{def:prgm}) and the execution of $\mathrm{i_{RV}}(P, M, S)$ (Definition~\ref{def:confsys}) composed of the same program  $P$, the debugger, a monitor $M$ (Definition~\ref{def:monitor}) and the empty scenario $S$ (Definition~\ref{def:scenario}). We denote by $\runPrgm$ the execution of $P$ that is a sequence of its configurations.
We prove a relation between the executions under the following conditions:
\begin{condition}
\label{cond1}
    The execution of the program terminates, i.e. it eventually reaches instruction $\stopInstr$.
\end{condition}

\begin{condition}
\label{cond2}
    The set of commands of the debugger that can be used is restricted to:
    \begin{enumerate}
        \item $\mathtt{load \; monitor}$ (run once, at the very beginning)
        \item $\mathtt{continue}\, n$
        \item $\mathtt{break} \; a$
        \item $\mathtt{watch} \, mode \; a, a \in \Address$
        \item $\mathtt{checkpoint}$
        \item $\mathtt{step}$
    \end{enumerate}
\end{condition}
Combined together, Conditions~\ref{cond1} and \ref{cond2} ensure that the execution is not altered.
Specifically, the user cannot restore any checkpoint (see \defref{def:checkpoint})
using the $\mathtt{restart}\, n$ and the command $\mathtt{exit}$ is used exactly once: when the execution of the program reaches the instruction $\stopInstr$ (the user cannot abort the execution of the program).

The relation between the program executions is defined as a relation between a configuration of the \irv-program and a configuration of the program.
\begin{definition}[Relation $\correspTo$ (corresponds to)]
    Let $\confIRV = (\inmathlist{(m_{p}', \pc'), (q_d, \Bp, \Wp, \Checkp), M, m_s})$ be a configuration of $\irv(P, M, S)$ and $ c_P = (m_p, \pc) $ a configuration of $P$.
    A configuration $\confIRV$ is said to correspond to a configuration $c_P$, denoted by $\confIRV \correspTo c_P $, if $\removeAllBPs(m_{p}', \Bp) = m_p $ and if $\pc' = \pc$.
\end{definition}
That is, $\confIRV$ corresponds to $c_P$ when the memory of the program in the system $m_{p}'$ is the same as the memory of the initial program $m_p$ when instrumentation is removed from $m_{p}'$ and the current address is the same in both configurations ($\pc' = \pc$).

In addition, we need a last condition on function $\runInstr$, the transition function that updates the state of the program.
\begin{condition}
    \label{assumption:runInstrNoInstrAccess}
    Function $\runInstr$ does not access to the executable portion of the memory (the instructions of the program); except for reading the cell at the current address ($\pc$).
\end{condition}
The evolution of system $\irv(P,M,S)$ is driven by functions implemented in the debbuger.
Some of these functions are used by the restricted set of commands of the debugger given in Conditions~\ref{cond1} and~\ref{cond2}.
These functions update (a part of) the state of the system.
Let $\F=\{\inmathlist{\instrument, \applyEvents, \normalStep, \setBP, \interactiveStep}\}$ be the set of these functions.
These functions can take (a part of) a configuration of the system and possibly other parameters, and return (a part of) a new configuration.
Let $\confIRV$ be a configuration of the system.
For any $f \in \F$, we allow the following (abusive) notation for readability: $\confIRV' = f(\confIRV, \dots)$, where $\confIRV'$ is the new configuration of the system after a call to $f$ and $\confIRV$ is the configuration of the system before the call.

The preservation of the executions of the initial program is stated in \theoref{theo}:
\begin{theorem}
\label{theo}
	Let $\confIRVinit$ and $c_{P}^{\mathrm{init}} = (m_p^0, \pc^0)$ be the initial configurations of $\irv(P,M,S)$ and $P$, respectively.
	Let $\confIRV$ and $c_P$ be configurations of $\irv(P,M,S)$ and $P$.
	Let $f \in \F$ be a function updating the configuration of $\irv(P,M,S)$.
	Under Conditions~\ref{cond1}, \ref{cond2}, and~\ref{assumption:runInstrNoInstrAccess}, the following assertions hold:
	\begin{enumerate}
		\item $\confIRVinit \correspTo c_{P}^{\mathrm{init}}$.
		\item If $\confIRV \correspTo c_P$, then $f(\confIRV) \correspTo c_P$ or $f(\confIRV) \correspTo \runInstr(c_P)$.
	\end{enumerate}
\end{theorem}
The first assertion holds true by construction: it is a direct consequence of \defref{def:confsys}.

To prove the second assertion, we shall use three intermediate predicates on the configurations of $\irv(P, M, S)$.

\begin{itemize}
    \item
	    Predicate $\instrOrigOrBPPredicate$ holds on a configuration $\confIRV$ when, in the memory, the instructions are either breakpoints or instruction of the original program:
	    \[
	    \forall a \in \InstrAddress(P), m_p^{\irv}[a] = \BREAK \lor m_p^{\irv}[a] =  m^0_p[a]
	    \]
	    ($\InstrAddress(P)$ is the set of addresses at which program instructions are found).
    \item
	    Predicate $\bpConsistentPredicate$ holds when, if there is a breakpoint instruction at a memory address, then the debugger is aware of this breakpoint, and the instruction associated to this breakpoint is the instruction in the orignal program:
	    \[
	    \forall a \in \InstrAddress(P), m_p^{\irv}[a] = \BREAK \implies \exists (a, \instr) \in \Bp\, : \, \instr = m^0_p[a].
	    \]
    \item
	    Predicate $\irvConfCorrespToPrgmPredicate$ holds when one can associate a configuration of the \irv-program to a configuration of the initial program:
	    \[
	    \exists c_P \in \runPrgm: \confIRV \correspTo c_P .
	    \]
\end{itemize}

These predicates are tools for proving the second point of \theoref{theo}.



The proof is organized as follow. We first prove that each relevant function in the pseudo code behaves correctly, that is, after the call to the function with a configuration that satisfies the three predicates, the returned configuration $\confIRV'$ also satisfies the three predicates and \pointtwo{\confIRV'}. We then prove that each command of the debugger in the restricted set described in Condition~\ref{cond2}, by using these functions, behaves correctly.

\hone{By function}

We first prove that functions called when the user issues a command in the debugger do not prevent the system from simulating the original program.

\htwo{Function $\setBP$}
\label{sec:proofsetbp}

Function $\setBP$ sets a breakpoint.

Let $\confIRV$ be the configuration of the system before the call to function $\setBP$ such that \R{\confIRV}.
Let $\confIRV'$ be the configuration of the system after the call to function $\setBP$.

A call to $\setBP$, defined in Section~\ref{sec:instr}, replaces an instruction in the memory of the program by an instruction $\BREAK$ at the given address $ a \in \InstrAddress(P) $ and adds a breakpoint in the breakpoint list containing $a$ and the instruction that is replaced.

Because we have $\instrOrigOrBPPredicate({\confIRV})$, this instruction is either $ \BREAK $ or the original instruction at address $ a $ in $ P $.
In the first case, $\bpConsistentPredicate(\confIRV)$ ensures that the original instruction is already kept in the list of breakpoints so $ \bpConsistentPredicate(\confIRV') $ is satisfied.
In the second case, the original instruction is stored and $\bpConsistentPredicate(\confIRV')$ is satisfied.
$\instrOrigOrBPPredicate(\confIRV')$ is also verified as the memory of the program has been modified so that at each address, the value is either the same as before or is $\BREAK$.

Because we have $\irvConfCorrespToPrgmPredicate(\confIRV)$, $\instrOrigOrBPPredicate(\confIRV')$, $\bpConsistentPredicate(\confIRV')$ are verified and values in the memory have not been modified, $\confIRV' \correspTo c_{P}$ and \pointtwo{\confIRV'}.

Since $ \forall \confIRV, \confIRV \correspTo c_P \implies \setBP(\confIRV) \correspTo c_P$ and \R{\setBP(\confIRV)}, The configuration obtained after several calls to $\setBP$ verifies point 2 of \theoref{theo}.

\htwo{Function $\instrument$}
\label{sec:proofinstrument}

Let $\confIRV=(\inmathlist{(m_{p}^{\irv}, \pc^{\irv}), (q_d, \Bp, \Wp, \Checkp),M, m_s})$ be the configuration of the system before the call to function $\instrument$ such that \R{\confIRV}.
Let $c_P $ be such that $\confIRV \correspTo c_P$ (possible because $\irvConfCorrespToPrgmPredicate(\confIRV)$).
Let $\confIRV'=(\inmathlist{(m_{p}^{\irv'}, \pc^{{\irv}'}), (q_d', \Bp', \Wp', \Checkp'),M', m_s'})$ be the configuration of the system after the call to function $\instrument$.
Function $\instrument$ makes one or more calls to $\setBP$ on Line~\ref{algoline:instrumentcalltosetbp}. \R{\confIRV'} and because $\confIRV' \correspTo c_P$, \pointtwo{\confIRV'}. See \secref{sec:proofsetbp}.

\htwo{Function $\normalStep$}
\label{sec:proofnormalstep}

Function $\normalStep$ executes one instruction of the program, when there is no breakpoint at the current instruction.

Let $\confIRV=(\inmathlist{(m_{p}^{\irv}, \pc^{\irv}), (q_d, \Bp, \Wp, \Checkp),M, m_s})$ be the configuration of the system before the call to function $\normalStep$ such that \R{\confIRV}.
Let $\confIRV'=(\inmathlist{(m_{p}^{\irv'}, \pc^{{\irv}'}), (q_d', \Bp', \Wp', \Checkp'),M', m_s'})$ be the configuration of the system after the call to function $\normalStep$.

Function $\runInstr$ is run, we obtain the new state: $\confIRV^{\mathrm{run}} = (\inmathlist{(m_{p}^{{\mathrm{run}}}, \pc^{{\mathrm{run}}}), (q_d, \Bp^{{\mathrm{run}}}, \Wp, \Checkp),M, m_s})$.
Because of Assumption~\ref{assumption:runInstrNoInstrAccess}, $\instrOrigOrBPPredicate({\confIRV^{\mathrm{run}}})$ and $\bpConsistentPredicate({\confIRV^{\mathrm{run}}})$ hold and because there exists $k \in \mathbb{N}$ such that $ \confIRV \correspTo c_P$, ${\confIRV^{\mathrm{run}}} \correspTo \runInstr(c_P)$.

Function $\handleWP$ (Line~\ref{algoline:normalstephandlewp}) either does not change the state of the system except for the state of the debugger which becomes interactive, so $\confIRV'=(\inmathlist{(m_p^{\irv}, \pc^{\irv}), (\Interactive, \Bp, \Wp, \Checkp),M, m_s}$, or calls $\applyEvents$ with state $ \confIRV $ (See \secref{sec:proofapplyevents}). In both cases, \R{\confIRV'} and \pointtwo{\confIRV'} (see \secref{sec:proofapplyevents}).

\htwo{Function $\applyEvents$}
\label{sec:proofapplyevents}

Function $\applyEvents$ removes all breakpoints from the memory~(1),
removes the breakpoints related to the current state of the property in the configuration of the debugger~(2),
applies before events that are related to the current breakpoint~(3),
runs the current instruction~(4),
applies after events that are related to the current breakpoint~(5) and
adds the instrumentation related to the new state of the property~(6).

Let $\confIRV=(\inmathlist{(m_{p}^{\irv}, \pc^{\irv}), (q_d, \Bp, \Wp, \Checkp),M, m_s})$ be the configuration of the system before the call to function $\applyEvents$ such that \R{\confIRV}.

Let $ c_P $ be such that $\confIRV \correspTo c_P$ (possible because $\irvConfCorrespToPrgmPredicate(\confIRV)$).
Let $\confIRV'=(\inmathlist{(m_{p}^{\irv'}, \pc^{{\irv}'}), (q_d', \Bp', \Wp', \Checkp'),M', m_s'})$ be the configuration of the system after the call to function $\applyEvents$.

\begin{enumerate}
    \item
	    First, all breakpoints are removed from $m_p^{\irv}$~(Line~\ref{algoline:aermallbp}).
	    Let $m_p^{{tmp,1}}$ be the result of this process and $\confIRV^{tmp,1} = (\inmathlist{(m_p^{{tmp,1}}, \pc^{\irv}), (q_d, \Bp, \Wp, \Checkp),M, m_s})$.
	    As there are no more addresses at which there is a $\BREAK$ instruction in $m_p^{{tmp,1}}$ , $\instrOrigOrBPPredicate({\confIRV^{tmp,1}})$ and $\bpConsistentPredicate({\confIRV^{tmp,1}})$ are verified.
	    ${\confIRV^{tmp,1}} \correspTo c_P$.
    \item
	    The function $\unInstrument$, defined in \algoref{algo:setbp}, is called~(Line~\ref{algoline:aermmonbp}).
	    This function removes some breakpoints from the list of breakpoints.
	    Let $B^{{tmp,2}}$ this new list of breakpoints and $\confIRV^{tmp,2} = (\inmathlist{(m_p^{{tmp,1}}, \pc^{\irv}), (q_d, B^{{tmp,2}}, \Wp, \Checkp),M, m_s})$.

		$\instrOrigOrBPPredicate({\confIRV^{tmp,2}})$ and $\bpConsistentPredicate({\confIRV^{tmp,2}})$ are verified and ${\confIRV^{tmp,2}} \correspTo c_P$.

	    As we assume an empty scenario, the memory and the list of breakpoints are not modified and the state remains unchanged.
    \item
	    The function $\runInstr$ is run~(Line~\ref{algoline:aeri}), we obtain the new state $\confIRV^{tmp,3} = (\inmathlist{(m_p^{{tmp,3}}, \pc^{{tmp,4}}), (q_d, B^{{tmp,2}}, \Wp, \Checkp),M, m_s})$.
	    Because of Assumption~\ref{assumption:runInstrNoInstrAccess}, $\instrOrigOrBPPredicate(\confIRV^{tmp,3})$ and $\bpConsistentPredicate(\confIRV^{tmp,3})$ hold and because $\confIRV^{tmp,2} \correspTo c_P$, $\confIRV^{tmp,3} \correspTo \runInstr(c_P)$.
	\item
		While applying after events~(Line~\ref{algoline:aeafter}), $\applyEvent$ updates the memory and the list of breakpoints only if a scenario element to be executed.
	    As we assume an empty scenario, the memory and the list of breakpoints are not modified.
	    By construction of $\restoreAllBPs$, the new memory $m_p^{{tmp,4}}$ contains instructions $\BREAK$ only at addresses that have corresponding breakpoints in the list of breakpoints. \R{\confIRV^{tmp,4}} with $\confIRV^{tmp,4}=(\inmathlist{(m_p^{{tmp,4}}, \pc^{{tmp,4}}), (q_d, B^{{tmp,2}}, \Wp, \Checkp),M, m_s})$.
	\item
		The function $\instrument$, is called on~Line \ref{algoline:aeinstr} with state $\confIRV^{tmp,4}$. The new state \R{\confIRV'} (see \secref{sec:proofinstrument}) and $\confIRV' \correspTo \runInstr(P)$, so \pointtwo{\confIRV'}.
\end{enumerate}

\htwo{Function $\handleStepBP$}
\label{sec:proofhandlestepbp}
Function $\handleStepBP$ handles the case when a step needs to be executed, but there is a breakpoint at the current instruction.

Let $\confIRV=(\inmathlist{(m_{p}^{\irv}, \pc^{\irv}), (q_d, \Bp, \Wp, \Checkp),M, m_s})$ be the configuration of the system before the call to function $\handleStepBP$ such that \R{\confIRV}.
Let $ c_P $ be such that $\confIRV \correspTo c_P$ (possible because $\irvConfCorrespToPrgmPredicate(\confIRV)$).
Let $\confIRV'=(\inmathlist{(m_{p}^{\irv'}, \pc^{{\irv}'}), (q_d', \Bp', \Wp', \Checkp'),M', m_s'})$ be the configuration of the system after the call to function $\handleStepBP$.

Function $\handleStepBP$ has two cases:

\begin{enumerate}
    \item
        A breakpoint is set by the monitor at the current address~(Line~\ref{algoline:handlestepmonbp}).

        Function $\applyEvents$, defined in \algoref{algo:applyEvent}, is called. See \secref{sec:proofapplyevents}.
    \item
        A breakpoint is not set by the monitor, but by the user, at the current address (Line~\ref{algoline:handlestepuserbp}.

        The breakpoint is replaced by the original instruction in the memory: $m_p^{{tmp,1}} = m_p^{\irv}[\pc^{\irv} \mapsto m_p^0[\pc^{\irv}]]$. Let $\confIRV^{tmp,1}=(\inmathlist{(m_p^{{tmp,1}}, \pc^{\irv}), (q_d, \Bp, \Wp, \Checkp),M, m_s})$ be the state of the system at this point.
        $\instrOrigOrBPPredicate({{tmp,1}})$ and $\bpConsistentPredicate({{tmp,1}})$ are verified because the set of addresses at which an instruction $\BREAK$ is present in this new memory ${m}_p^{{tmp,1}}$ is the set of addresses at which an instruction $\BREAK$ is present in $m_p^{{tmp,1}}$, without $\pc^{\irv}$.
        Because values in $m_p^{{tmp,1}}$ are the same as in $m_p^{{tmp,1}}$, $\confIRV^{tmp,1} \correspTo c_P$.

		A normal execution step is performed with the function $\textsc{normalStep}$ with state $\confIRV^{tmp,1}$. The new state \R{\confIRV^{tmp,2}} and ${\confIRV^{tmp,2}} \correspTo c_P$ (see \secref{sec:proofnormalstep}).

		The returned state $\confIRV'$ is the state $\confIRV^{tmp,2}$ in which a breakpoint instruction is added at the initial pc in the program memory: $ \confIRV' = (\inmathlist{(m_{p}^{{tmp,2}}[\pc^{\irv} \mapsto \BREAK], \pc^{{tmp,2}}), (q_d^{{tmp,2}}, \Bp^{{tmp,2}}, \Wp^{{tmp,2}}, \Checkp^{{tmp,2}}),M^{{tmp,2}}, m_s^{{tmp,2}}}) $.
        As the set of addresses at which an instruction $\BREAK$ is present in the program memory of state $\confIRV'$ is the same as the set of addresses at which an instruction $\BREAK$ is present in $\confIRV^{tmp,2}$, $\instrOrigOrBPPredicate(\confIRV')$ and $\bpConsistentPredicate(\confIRV')$ are verified. Because $\confIRV^{tmp,2} \correspTo \runInstr(c_P)$, $\confIRV' \correspTo P_{k+1}$ so $\irvConfCorrespToPrgmPredicate(\confIRV')$ is verified and \pointtwo{\confIRV'}.
\end{enumerate}

\htwo{Function $\interactiveStep$}
\label{sec:proofinteractivestep}

Function $\interactiveStep$, defined in \algoref{algo:step}, executes one step in the execution of the program
There are two cases:

\begin{enumerate}
    \item
        The current address in the program is a breakpoint~($m_p^{\irv}[\pc^{\irv}] = \BREAK$, Line~\ref{algoline:stepbystepbreak}).
        Function $\handleStepBP$ updates the state of the system. See section \secref{sec:proofhandlestepbp}.

    \item
        The current address in the program is not a breakpoint ($m_p^{\irv}[\pc^{\irv}] = \BREAK$). A normal execution step is performed with the function $\textsc{normalStep}$. See section \secref{sec:proofnormalstep}.
\end{enumerate}

\hone{Commands of the Debugger}

%

We now prove that each command the user can issue does not prevent the system from simulating the program.
The behavior of the system for each command is described in \algoref{algo:general}.

\htwo{The user issues the command $\mathtt{load \; monitor}.$}

When this command is issued in \algoref{algo:interactive} on Line~\ref{algoline:loadmon}, the memory of the program and the list of breakpoints are modified by the function $\instrument$ defined in \algoref{algo:setbp}.
See \secref{sec:proofinstrument}.

\htwo{The user issues the command $\mathtt{continue}.$}

When the user issues the command $\mathtt{continue}$ in \algoref{algo:interactive} on Line~\ref{algoline:continue}, the function $\interactiveStep$ is called.
See \secref{sec:proofinteractivestep}.
The mode of the debugger becomes $\Passive$, making the system run function $\normalStep$ a number of times (see \secref{sec:proofnormalstep}), and run function $\handleBP$ if a breakpoint is encountered. Function $\handleBP$ either does not modify the memory of the program and the list of breakpoints, or calls the function $\applyEvents$. See \secref{sec:proofapplyevents}.

\htwo{$\mathtt{break}\, a$}

When the user issues the command $\mathtt{break}$ in \algoref{algo:interactive} on Line~\ref{algoline:setbreak}, the function $\setBP$. See \secref{sec:proofsetbp}.

\htwo{$\mathtt{step}$}

When the user issues the command $\mathtt{continue}$ in \algoref{algo:interactive} on Line~\ref{algoline:continue}, the function $\interactiveStep$ is called.
See \secref{step:proofinteractivestep}.

\htwo{$\mathtt{watch}$, $\mathtt{checkpoint}$ or the user interrupts the execution}

These commands~(in \algoref{algo:interactive}, Lines~\ref{algoline:setwatch}, \ref{algoline:setcheck} and \ref{algoline:interrupt}) do not affect the list of breakpoints nor the memory of the program, so simulation is unaffected.

\htwo{Normal execution}

During a normal execution~(\algoref{algo:general}, Line~\ref{algoline:normalexec}), the function $\textsc{normalStep}$ is called. See \secref{sec:proofnormalstep}.

\htwo{A breakpoint is reached}

When a breakpoint is reached~(\algoref{algo:general}, Line~\ref{algoline:dbginstr}), function $\handleBP$ is called.
This function either does not modify the memory of the program and the list of breakpoints, or calls the function $\applyEvents$. See \secref{sec:proofapplyevents}.
\fi

\hzero{Implementation: \ce}
\label{chap:implementation}
%
%
To validate and evaluate \irv{} in terms of usefulness and performance, we implemented it in a tool called \ce{}.
We overview \ce{}\ifdefined \BLINDVERSION\footnote{this is the anonymized name of the tool.}\fi{ }and give some details about its architecture in \secref{secpoc}.
In \secref{sec:syntax}, we describe the syntax used in \ce{} to write properties.
\ifdefined \BLINDVERSION \else
We explain how to use \ce{} in \secref{sec:usage}.
\fi

\hone{Overview}
\label{secpoc}

\begin{figure}[t]
	\ifdefined \BLINDVERSION
		\centerline{\includegraphics[width=0.9\linewidth]{archi-anon.pdf}}
	\else
		\centerline{\includegraphics[width=0.9\linewidth]{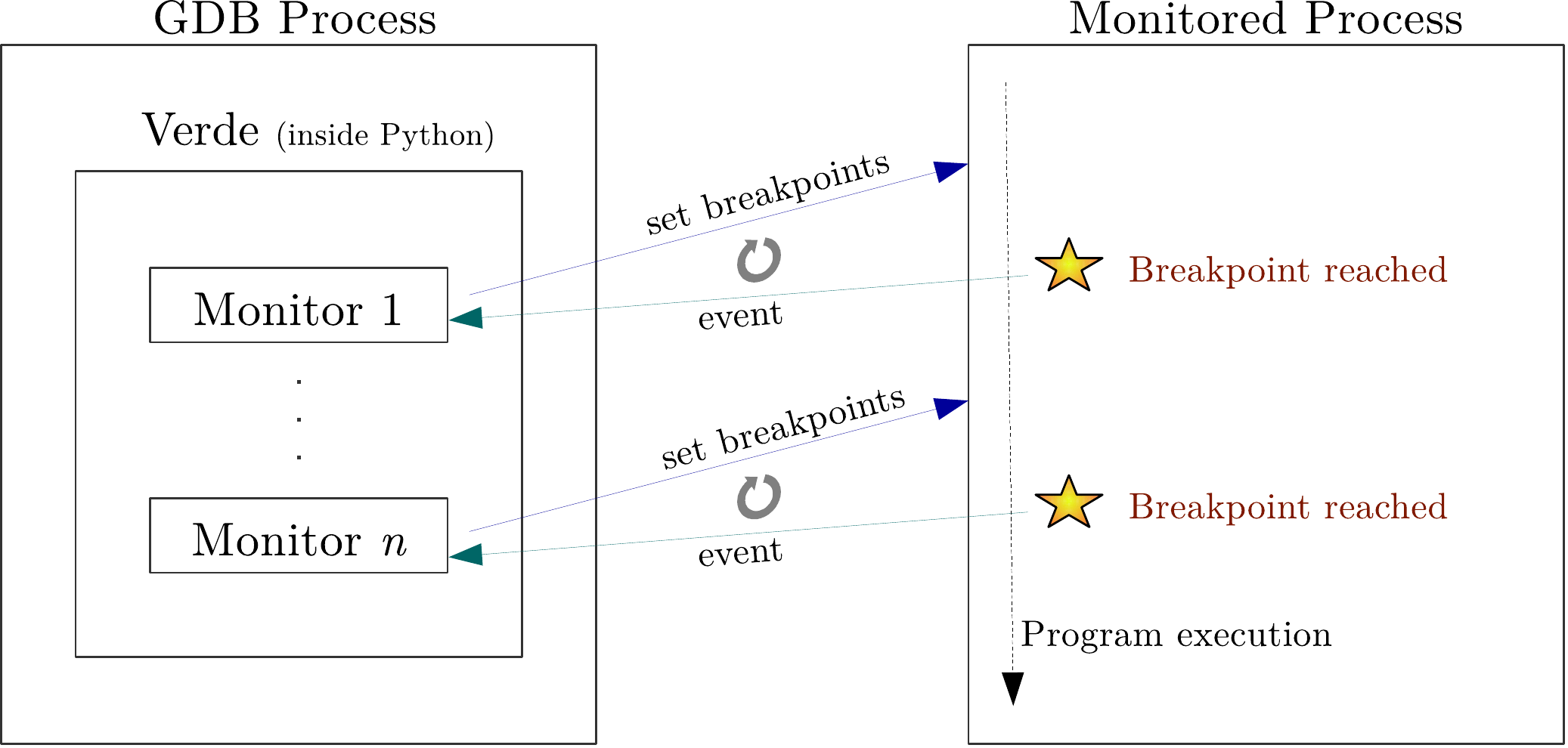}}
	\fi
	\vspace{-1em}
	\caption{Instrumentation in \ce{}}
	\label{fig:archi}
\end{figure}

\ifdefined \RLYSHORTVERSION
\begin{figure}
	\begin{center}
		\includegraphics[scale=0.33]{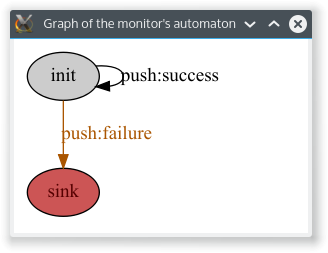}
	\end{center}
	\vspace{-1em}
	\caption{Graphical view the property given in
		Figure~\ref{fig:checkprop}.}
	\label{fig:showgraph}
\end{figure}
\else
\begin{figure}
	\begin{center}
		\includegraphics[scale=0.5]{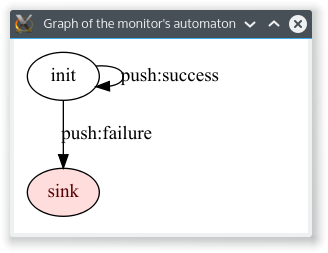}
		\includegraphics[scale=0.5]{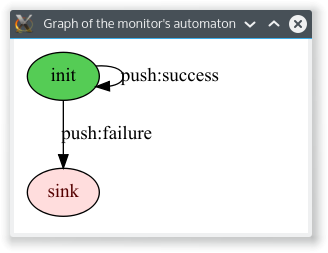}
		\includegraphics[scale=0.5]{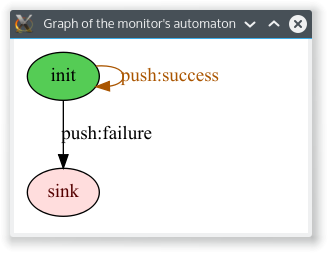}
		\includegraphics[scale=0.5]{graphsink.png}
	\end{center}
	\caption{During the execution of the property given in
		Figure~\ref{fig:checkprop}, the following graphs can be seen respectively
		before initialization of the property initialization, on initialization, while
		the property is verified and when the property becomes falsified. Light red, red,
		brown and gray respectively correspond to non accepting state, a
		current non accepting state, a transition taken during the last state change
		and a state which was current before the last state change. Graphs
		are automatically drawn using Graphviz and colors animated during the
		execution.}
	\label{fig:showgraph}
\end{figure}
\fi

\Ce{}\ifdefined \BLINDVERSION \else\footnote{\Ce{} can be downloaded at \url{https://gitlab.inria.fr/monitoring/verde}.}\fi{ }is written in Python and works seamlessly as a GDB plugin by extending GDB Python interface.
\Ce{} can be used with any program written in a programming language supported by GDB.
\Ce{} supports the verification of several properties by means of monitors working independently.
Each monitor sets and deletes breakpoints according to the events that are relevant to its current state.
\Ce{} provides a graphical and animated view of the properties being checked at runtime.
The view facilitates understanding the current evaluation of the property and, as a consequence, the program.
\Ce{} also lets the developer control the monitors and access their internal state (property instances, current states, environments).
\Ce{} is called by GDB when GDB handles breakpoints in the monitored program.
When a breakpoint is reached, the state of the property is updated and the execution is
resumed.
Figure~\ref{fig:archi} depicts the execution of a program with \ce{}.

\ifdefined \RLYSHORTVERSION \else{
\hone{Architecture of \Ce{}}
\label{sec:codeorga}
\ifdefined \RLYSHORTVERSION \else{
\begin{figure*}
	\centering
	\includegraphics[width=1\linewidth]{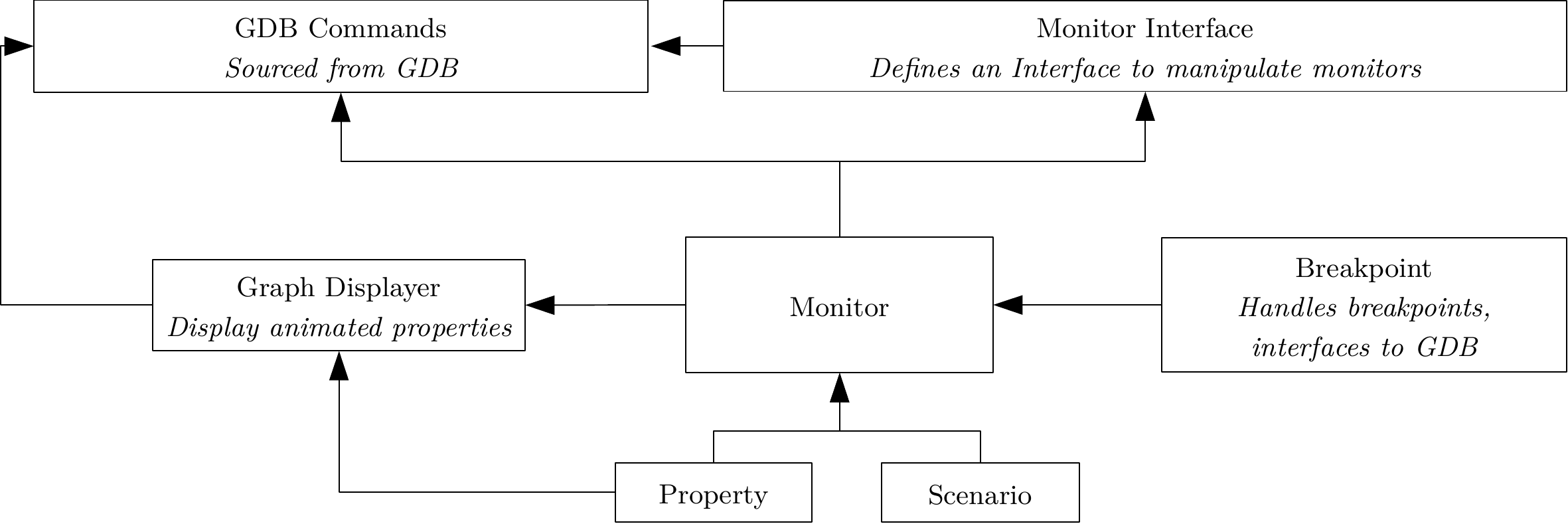}
	\caption{Organization of the code of \ce{}}
	\label{fig:check-exec-orga}
\end{figure*}

The organization of the code is depicted in Figure~\ref{fig:check-exec-orga}.
\fi
The central part of \ce{} is the \texttt{Monitor} class.
A \texttt{Monitor} object is instantiated for each property checked at runtime.
When the developer's properties and scenarios are read from files, the result is
stored in instances of classes \texttt{Property} and \texttt{Scenario}.
These instances are used to build monitors.

\ce{} and its monitors are controlled with the GDB command line interface using
commands defined in the module \texttt{GDB Commands}.
This module defines the interface between the GDB user and \ce{}.
Theses commands expose a part of the interface defined in the module \texttt{Monitor Interface}.
This interface is meant to remain a stable interface to access monitors.
It does not give access to the internal structures that are exposed by the Monitor class and that are not relevant for the end user.
\ifdefined \SHORTVERSION
This interface is given in the extended version of this paper.
\else
This interface is given in Appendix~\ref{apdx:cemi}.
\fi
Module \texttt{Breakpoint} defines the interface between the monitors and GDB.
It defines debugger-independent methods to handle breakpoints.
Module \texttt{Graph Displayer} defines the graphical view of running monitors.
If the view is enabled, \ce{} shows the property as a graph using Graphviz.
As the current state of the monitor changes, the graphical view is updated: the current state is shown in green if it is accepting, in red if it is not accepting.
Taken transitions are represented in brown.
Module \texttt{Property} handles the property model used in \ce{} and trace slicing.

\ifdefined \RLYSHORTVERSION
\else
\begin{figure}
\begin{lstlisting}[basicstyle=\footnotesize,escapeinside={(@}{@)},literate=
	{[}{{\textcolor{brown}{\big [}}}1
	{]}{{\textcolor{brown}{\big ]}}}1
	{*}{{\textcolor{blue}{*}}}2
	{|}{{\textcolor{blue}{|}}}2
	{+}{{\textcolor{blue}{+}}}2
]

[slice on [(@\textit{param}@),]+](@{\textcolor{blue}{?}}@)

[initialization {
    (@\textit{Python code}@)
}](@{\textcolor{blue}{?}}@)

[state(@
@) (@\textit{state\_name}@)[[non-](@{\textcolor{blue}{?}}@)accepting](@{\textcolor{blue}{?}}@) [(@\textit{action\_name}@)()](@{\textcolor{blue}{?}}@) {
  [transition {
    [before|after](@{\textcolor{blue}{?}}@) event (@\textit{event\_name}@)([(@\textit{param}@),]*) [{
      (@\textit{Python code returning}@) True False (@\textit{or}@) None
    }](@{\textcolor{blue}{?}}@)
    [success [{
        (@\textit{Python code}@)
    }](@{\textcolor{blue}{?}}@) [(@\textit{action\_name}@)()](@{\textcolor{blue}{?}}@) (@\textit{state\_name}@)](@{\textcolor{blue}{?}}@)
    [failure [{
        (@\textit{Python code}@)
    }](@{\textcolor{blue}{?}}@) [(@\textit{action\_name}@)()](@{\textcolor{blue}{?}}@) (@\textit{state\_name}@)](@{\textcolor{blue}{?}}@)
  }]*
}]+
\end{lstlisting}
\caption{Informal grammar for the automaton-based property description language
in \ce{}}
\label{fig:informalgr}
\end{figure}
\fi
\fi
\hone{Syntax of Properties}
\label{sec:syntax}
\begin{figure}[t]
\ifdefined\SHORTVERSION
\begin{lstlisting}[style=CheckExecStyle,deletekeywords={max},escapeinside={\%*}{*)}]
%*{\textcolor{blue}{slice on}} queue
*)
initialization: {N = 0; maxSize = 0}
state init accepting:
	transition:
		event queue_new(queue, size : int) { maxSize = size - 1 }
		success queue_ready
state queue_ready accepting:
	transition:
		event queue_push(queue, prod_id) { return N < maxSize }
		success { N = N + 1; print("nb elem: "+str(N));} queue_ready
		failure { print("%d made %d overflow!"% (prod_id, queue)) } sink
	transition:
		event queue_pop(queue, prod_id) { return N > 0 }
		success { N = N - 1; print("nb elem: "+str(N)) } queue_ready
		failure sink
state sink non-accepting sink_reached()
\end{lstlisting}
\else
\begin{lstlisting}[style=CheckExecStyle,deletekeywords={max},escapeinside={\%*}{*)}]
%*{\textcolor{blue}{slice on}} queue
*)
initialization {
	N = 0
	maxSize = 0
}

state init accepting {
	transition {
		event queue_new(queue, size : int) {
			maxSize = size - 1
		}
		success queue_ready
	}
}

state queue_ready accepting {
	transition {
		event queue_push(queue, prod_id) {
			return N < maxSize
		}

		success {
			N = N + 1
			print("nb elem: "+str(N))
		} queue_ready

		failure {
			print("%d made %d overflow!"% (prod_id, queue))
		} sink
	}

	transition {
		event queue_pop(queue, prod_id) {
			return N > 0
		}

		success {
			N = N - 1
			print("nb elem: "+str(N))
		} queue_ready

		failure sink
	}
}

state sink non-accepting sink_reached()
	\end{lstlisting}
\fi
	\vspace{-1em}
	\caption{\ce{} version of the property in Figure~\ref{fig:exautomaton}}
	\label{fig:checkprop}
\end{figure}

\Ce{} provides a custom syntax for writing properties in the model presented in \secref{sec:model} with slight modifications to allow more conciseness\footnote{We did not use pre-existing syntax in order to allow us flexibility as we experiment. Interfacing with existing monitoring tool is planned.}.\hideInRLYShortVersion{
An informal grammar is given in \figref{fig:informalgr}.
}\figref{fig:checkprop} gives a property used to check whether an overflow happens in a multi-threaded producer-consumer program.

First, the optional keyword \texttt{slice on} gives the list of slicing parameters.
Then, an optional Python code block initializes the environment of the monitor.
Then, states are listed, including the mandatory state \texttt{init}.
A state has a name, an optional annotation indicating whether it is accepting, an optional action name attached to the state and its transitions.
Transitions can be written with two destination states: a success (resp. failure)
state used when the guard returns \textsc{success} (resp. failure).
The guard can also return \textsc{not relevant} meaning that the transition will not be taken.
Each transition comprises the monitored event, the parameters of the event used
in the guard, the guard (optional), the success block and the failure block
(optional).
Success and failure blocks comprise an optional Python code block, an optional action name and the name of a destination state.
The guard is Python block code without any side effect that returns True (resp. False) if the guard succeeds (resp. fails) and None if the transition should not be taken.

\hone{Checkpointing}

\Ce{} features two process checkpointing techniques on Linux-based systems.
The first uses the native \texttt{checkpoint} command  of GDB.
This method is based on \texttt{fork()} to save the program state in a new process, which is efficient, as \texttt{fork} is implemented using Copy on Write.
A major drawback of this technique is that multithreaded programming is not supported since \texttt{fork()} keeps only one thread in the new process.
The second technique uses CRIU\footnote{Checkpoint/Restore In Userspace}, which supports multithreaded processes and trees of processes.
CRIU uses the \texttt{ptrace} API to attach the (tree of) process to be checkpointed and saves its state in a set of files.
CRIU supports incremental checkpointing by computing a differential between an existing checkpoint and a checkpoint to create.
It can make the system track memory changes in the process to speed this computation.

\ifdefined \SHORTVERSION \else
\fi

\ifdefined \BLINDVERSION \else
\hone{Using \Ce{}}
\label{sec:usage}
A typical usage session begins by launching \texttt{gdb} and \ce{} (which can be automatically loaded by configuring GDB appropriately).
Then, the user loads one or several properties.
Additional python functions, used in properties, can be loaded at the same time.
A scenario can also be loaded.
Then, the user starts the execution.
It is also possible to display the graph of the property with the \texttt{show-graph}
subcommand (see \figref{fig:showgraph}).

{\footnotesize{
\begin{lstlisting}
$ gdb ./my-application
(gdb) verde load-property correct-behavior.prop \
      functions.py
(gdb) verde load-scenario default-scenario.sc
(gdb) verde show-graph
(gdb) verde run-with-program
...
[verde] Initialization: N = 0
[verde] Current state: init (N = 0)
queue.c: push!
[verde] Current state: init
...
queue.c: push!
[verde] GUARD: nb push: 63
[verde] Overflow detected!
[verde] Current state: sink (N = 63)
[Execution stopped.]
(gdb)
\end{lstlisting}
}}

\hideInRLYShortVersion{
\Ce{} provides more fine-tuned commands to handle cases when
properties and functions need to be loaded separately, or when properties and the
program need to be run at different times.
\ifdefined \SHORTVERSION
A list of commands is given in the extended version of this paper.
\else
A list of commands is given in Appendix~\ref{apdx:cecmdlist}
\fi
}

\fi

\hzero{Evaluation}
\label{sec:eval}
%
We report on six experiments carried out with \ce{} to measure its usefulness in finding and correcting bugs and its efficiency from a performance point of view\ifdefined\BLINDVERSION \else \footnote{A video and the source codes needed for reproducing the benchmarks are available at \url{http://gitlab.inria.fr/monitoring/verde}.}\fi.
We discuss the objective and possible limitations (threat to validity) of each experiment.
These experiments also illustrate how a developer uses \ce{} in practice.
%
%
\hone{Correcting a Bug in \zsh{}}
\label{subsec:evalzsh}
%
In \zsh{}, a widely-used UNIX shell, a segmentation fault happened
when trying to auto-complete some inputs like \texttt{!> .} by hitting the tab key right after character $\mathtt{>}$.
We ran \zsh{} in GDB, triggered the bug and asked for a backtrace that
leads to a long and complicated function, \texttt{get\_comp\_string},
calling another function with a null parameter \texttt{itype\_end}, and then making \zsh{} crash.
Instead of trying to read and understand the code or doing step by step debugging
from the beginning of this function, we wrote a property to monitor the writes to the variable  passed to function~\texttt{itype\_end} and a scenario that prints the backtrace each time the state of the property changes.
This lets us see that the last write to this variable nulls it.
We were able to prevent the crash by adding a null check before a piece of code that seems to assume that the variable is not null and that contains the call to \texttt{itype\_end} that lead to the crash\footnote{\ifdefined\SHORTVERSION \else The code of the property is in Appendix~\ref{apdx:zshprop}. \fi
We worked on commit 85ba685 of \zsh{}.}.
While \ce{} was not used to discover the bug\footnote{The bug was reported at \url{https://sourceforge.net/p/zsh/bugs/87/}}, it helped us determining its source in the code of \zsh{} and fixing it. A fix has since been released.
%
\hone{Multi-Threaded Producer-Consumers}
\label{subsec:evalprodcons}
%
This experiment is purposed to check whether our approach is realistic in terms of usability.
We considered the following use-case: a developer works on a multi-threaded application in which a queue is filled by 5 threads and emptied by 20 threads and a segmentation fault happens in several cases.
We wrote a program deliberately introducing a synchronization error, as well as a property (see \figref{fig:exautomaton}) on the number of additions in a queue in order to detect an overflow.
The size of the queue is a parameter of the event \texttt{queue\_new}.
The function \texttt{push} adds an element into the queue.
A call to this function is awaited by the transition defined at line~15 of \figref{fig:checkprop}.
We ran the program with \ce{}.
The execution stopped in the state \texttt{sink}~(defined at line~39 of \figref{fig:checkprop}).
In the debugger, we had access to the precise line in the source code from which the function is called, as well as the complete call stack.
Under certain conditions (that we artificially triggered), a mutex was not locked, resulting in a queue overflow.
After fixing this, the program behaved properly.
In this experiment, we intentionally introduced a bug (and thus already knew its location).
However this experiments validates the usefulness of \ce{} in helping the programmer locate the bug: the moment the verdict given by the monitor becomes false can correspond to the exact place the error is located in the code of the misbehaving program.
%
\hone{Micro-benchmark}
\label{secmicrobench}
%
\begin{figure}[t]
\ifdefined \BLINDVERSION
	\centerline{\includegraphics[width=1\linewidth]{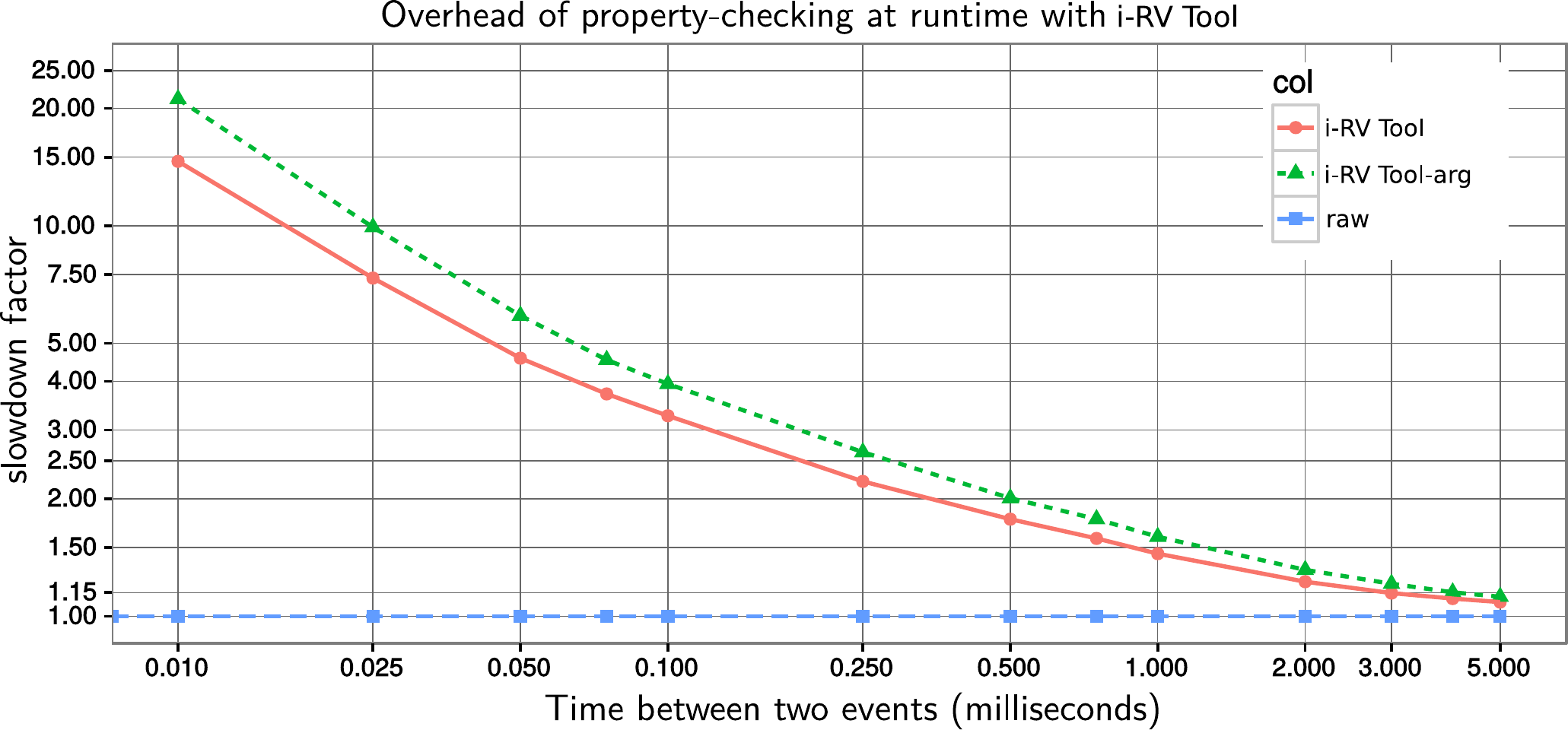}}
\else
	\centerline{\includegraphics[width=1\linewidth]{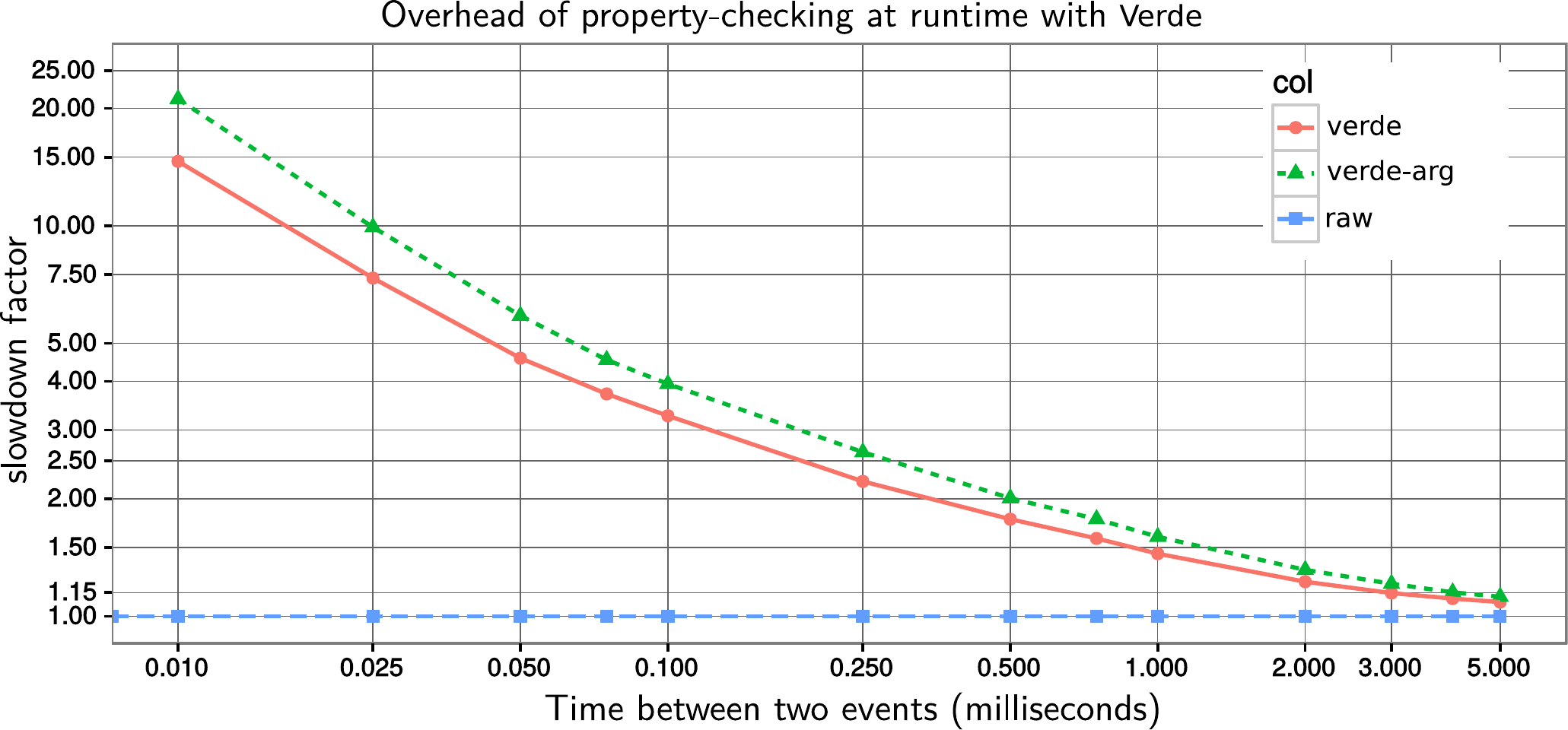}}
\fi
\vspace{-1em}
 \caption{Instrumentation overhead with \ce{}.}
\label{fig:bench}
\end{figure}

In this experiment, we evaluated the overhead of the instrumentation in function of the temporal gap between events.
We wrote a C program calling a NOP function in a loop.
To measure the minimal gap between two monitored events for which the
overhead is acceptable, we simulated this gap by a loop of a
configurable duration.
The results of this benchmark using a
Core i7-3770 @ 3.40~GHz (with a quantum time (process time slice) around 20 ms), under Ubuntu 14.04 and Linux 3.13.0, are
presented in \figref{fig:bench}.
The curve \texttt{\ifdefined \BLINDVERSION i-RV Tool-arg \else verde-arg \fi} corresponds to the evaluation of a property
which retrieves an argument from calls to the monitored function.
With 0.5~ms between two events, we measured a slowdown factor of 2.
Under 0.5~ms, the overhead can be significant.
From 3~ms, the slowdown is under 20~\% and from 10~ms, the slowdown is under~5~\%.
We noticed that the overhead is dominated by breakpoint hits.
The absolute overhead by monitored event, in the manner of the overhead of an argument retrieval, is constant.
We measured the mean cost of encountering a breakpoint during the execution.
We obtained 95~\textmu s on the same machine and around  300~\textmu s on a slower machine (i3-4030U CPU @ 1.90~GHz).
While this experiment does not give a realistic measure of the overhead added by the instrumentation, it is still useful to estimate the overhead in more realistic scenarios.
%
\hone{User-Perceived Performance Impact}
\label{sec:video}
%
\myparag{Multimedia Players and Video Games}
We evaluated our approach on widespread
multimedia applications: the VLC and MPlayer video players and the SuperTux 2D
platform video game. A property made the monitor set a breakpoint on the
function that draws each frame to the screen for these applications, respectively
\textit{ThreadDisplayPicture}, \textit{update\_video} and
\textit{DrawingContext::do\_drawing}.
For SuperTux, the function was called around 60 times per second.
For the video players, it was called 24 times per second. In each case, the number of frames per second was not affected
and the CPU usage remains moderated: we got an overhead of less than 10~\%
for the GDB process. These results correspond to our measurements in
\secref{secmicrobench}: there is a gap of 16~ms between two function calls
which is executed 60 times per second. Thus, our approach does not lead to a
significant overhead for multimedia applications when the events occur at fixed frequency.

\myparag{Opening and Closing Files, Iterators}
We evaluated the user-perceived overhead with widespread applications.
We ensured that all open files are closed with the Dolphin file manager, the
NetSurf Web browser, the Kate text editor and the Gimp image editor. Despite
some slowdowns, caused by frequent disk accesses, they remained usable.
Likewise, we checked that no iterator over hash tables of the GLib
library (\texttt{GHashTableIter}) that is invalidated was used. Simplest
applications like the Gnome calculator remained usable but strong slowdowns were
observed during the evaluation of this property, even for mere mouse movements.
In \secref{chap:future}, we present possible ways to mitigate these limitations.
\hideInShortVersion{
%
\hone{Dynamic Instrumentation on a Stack}
%
We measured the effects of the dynamic instrumentation on the performance.
A program adds and removes, alternatively, the first 100 natural integers in a stack. We checked that the integer 42 is taken out of the stack after being added.
A first version of this property leverage the dynamic instrumentation.
With this version, the call to the remove function was watched only when the monitor knew that 42 is in the stack. A second
version of the property made the monitor watch every event unconditionally.
The execution was 2.2 times faster than with the first version.
While this experiment used artificial properties, it shows that dynamic instrumentation has a positive impact on the overhead in that it improves performance.
}

\ifdefined\SHORTVERSION\else
%
\hone{Performance Impact of Checkpointing}
%
In this experiment, we measured the cost of checkpointing a process running in GDB with CRIU.
We wrote a C program that allocates an amount of memory given in parameter.
We set a checkpoint and restarted it ten times for different sizes, once using the memory as storage for checkpoints, once using a regular hard drive (see Table~\ref{table:checkrest}).
We noticed that checkpointing leads to acceptable costs for debugging purposes, ranging from 0.02 seconds for a 1 MB process to 0.3 seconds for a 1 GB process when storing checkpoints in RAM. We also noticed that the impact of checkpointing and restoring when using a hard drive as storage for saved checkpoints, as compared to a storage in RAM, is higher but remains in the same order of magnitude\footnote{This is probably due to caching mechanisms provided by the operating system.}.
The incremental checkpointing feature of CRIU improves performance when checkpointing a process several times.

\begin{table}
{\footnotesize
\begin{tabular}{|r|c|c|c|c|c|c|c|c|}
\cline{2-9}
\multicolumn{1}{c|}{} & \multicolumn{4}{c|}{RAM} &\multicolumn{4}{c|}{HDD} \\
\cline{2-9}
\multicolumn{1}{c|}{} & \multicolumn{2}{c|}{Checkpoint} & \multicolumn{2}{c|}{Restore} & \multicolumn{2}{c|}{Checkpoint} & \multicolumn{2}{c|}{Restore} \\
\hline
Process size (MB) & Avg (s) & ±   &  Avg (s) & ± & Avg (s) & ±   & Avg (s) & ± \\
\hline
1    &   0.0205 & 0.0026  & 0.0469 & 0.0134   & 0.0215 & 0.0048  & 0.0441 & 0.0129 \\
5    &   0.0243 & 0.0148  & 0.0490 & 0.0216   & 0.0238 & 0.0033  & 0.0447 & 0.0220 \\
10   &   0.0237 & 0.0052  & 0.0438 & 0.0106   & 0.0250 & 0.0021  & 0.0527 & 0.0260 \\
25   &   0.0274 & 0.0008  & 0.0524 & 0.0230   & 0.0337 & 0.0167  & 0.0505 & 0.0141 \\
50   &   0.0348 & 0.0046  & 0.0556 & 0.0107   & 0.0435 & 0.0031  & 0.0593 & 0.0151 \\
75   &   0.0433 & 0.0057  & 0.0678 & 0.0267   & 0.0545 & 0.0054  & 0.0626 & 0.0117 \\
100  &   0.0494 & 0.0067  & 0.0766 & 0.0173   & 0.0674 & 0.0101  & 0.0685 & 0.0123 \\
250  &   0.0918 & 0.0063  & 0.1086 & 0.0216   & 0.1370 & 0.0236  & 0.1101 & 0.0229 \\
500  &   0.1732 & 0.0688  & 0.1796 & 0.0191   & 0.2508 & 0.0475  & 0.1716 & 0.0107 \\
750  &   0.2347 & 0.0197  & 0.2454 & 0.0220   & 0.3645 & 0.0749  & 0.2360 & 0.0206 \\
1000 &   0.3018 & 0.0132  & 0.3098 & 0.0805   & 0.4839 & 0.1323  & 0.3018 & 0.0431 \\
\hline
\end{tabular}
}
\caption{Average time to checkpoint and restore in function of the size of the process, when checkpoints are saved on a Hard Disk Drive or in RAM.}
\label{table:checkrest}
\end{table}
\fi
%
\hone{Automatic Checkpointing to Debug a Sudoku Solver}
%
We evaluated i-RV by mutating the code of a backtracking Sudoku solver\footnote{\url{https://github.com/jakub-m/sudoku-solver}}.
This experiment illustrates the use of scenarios to automatically set checkpoints and add instrumentation at relevant points of the execution.
Sudoku is a game where the player fills a 9x9 board such that each row, each column and each 3x3 box contains every number between 1 and 9.
The solver reads a board with some already filled cells and prints the resulting board.
During the execution, several instances of the board are created and unsolvable instances are discarded.
We wrote a property describing its expected global behavior after skimming the structure of the code, ignoring its internal details.
No values should be written on a board deemed unsolvable or that break the rules of Sudoku (putting two same numbers in a row, a column or a box).
Loading a valid board should succeed.
We then wrote a scenario that creates checkpoints whenever the property enters an accepting state.
Entering a non-accepting state makes the scenario restore the last checkpoint and add watchpoints on each cells of the concerned board instance.
When watchpoints are reached, checkpoints are set, allowing us to get a more fine-grained view of the execution close to the misbehavior and choose the moment of the execution we want to debug.
This scenario allows a first execution that is not slowed down by heavy instrumentation, and precise instrumentation for a relevant part of it.
The solver is bundled with several example boards that it solves correctly.
We mutated its code using \texttt{mutate.py}\footnote{\url{https://github.com/arun-babu/mutate.py}} to artificially introduce a bug without us knowing where the change is.
When ran, the mutated program outputs "bad board".
We ran it with \irv. The property enters the state \texttt{failure\_load}.
When restoring a checkpoint and running the code step by step in the function that loads a board, the execution seems correct.
The code first runs one loop reading the board using \texttt{scanf} by chunks of 9 cells, and then a second loop iterates over the 81 cells to convert them to the representation used by the solver.
Setting breakpoints and displaying values during the first loop exhibits a seemingly correct behavior.
During the second loop, the last line of the board holds incorrect values.
Since we observed correct behavior for the first loop and the 72 first iterations of the second loop, and since both loops do not access the board in the same way, we suspected a problem with the array containing the board.
We checked the code and saw that the mutation happened in the type definition of the board, giving it 10 cells by line instead of 9.
A caveat of this experiment is that we had to choose the mutated version of the code such that the code violates the property.
We also introduced a bug artificially rather than working on a bug produced by a human.
However, the example can be generalized and illustrates how scenarios can be used for other programs, where checkpoints are set on a regular basis and execution is restarted from the last one and heavy instrumentation like watchpoints is used, restricting slowness to a small part of the execution.

\hzero{Related Work}
\label{sec:existing}
\irv{} is related to several families of techniques for finding and fixing bugs.
\myparag{Interactive and reverse debugging}
Tools used in interactive debugging are mainly debuggers such as GDB, LLDB and the Visual Studio debugger.
\hideInShortVersion{
GDB is a cross-platform free debugger from the Free
Software Foundation. LLDB is the cross-platform free debugger of the LLVM project,
started by the University of Illinois, now backed by various firms like Apple
and Google. The Visual Studio debugger is Microsoft's debugger.
}
Reverse debugging~\cite{engblom2012review,DBLP:conf/euc/GeorgievM14,DBLP:conf/pdp/RoosCM93} is a complementary debugging technique.
A first execution of the program showing the bug is recorded.
Then, the execution can be replayed and reversed in a deterministic way, guaranteeing that the bug is observed and the same behavior is reproduced in each replay.
UndoDB and rr are GDB-based tools allowing record and replay and reverse debugging with a small overhead.
i-RV also allows to restore the execution in a previous state using checkpoints, with the help of the monitor and the scenario, adding a level of automation.

\hideInShortVersion{
\myparag{Manual testing} Beta testing is a widespread way of testing. Most
obvious bugs can easily be spotted this way during the development of the
software. Modifications to the code are manually tested, possibly by a team
responsible for testing the software~\cite{itkonen2009testers}. Bugs are also
spotted by final users of the software, which, depending on the development
model, the kind of software and the severity of the bug, is more or less
undesirable.
\myparag{Automatic testing} Unit tests ensure that already-fixed bugs do not show again, to limit regressions
and to check the correctness of the code for a restricted set of
inputs~\cite{itkonen2007defect}.
Unit testing is a way to apply automatic testing.
Many unit testing frameworks exist.
JUnit and CppUnit are two examples of such frameworks.
Some research efforts have been carried out on
the automatic generation of unit tests. For
instance,~\cite{DBLP:conf/ecoop/CheonL02} defines a way to generate test oracles
from formal specifications of the expected behavior of a Java method or class.
}

\hideInShortVersion{
\myparag{Debugging}

A debugger has been written to type check program written
in~C~\cite{typecheckdbg} by tagging memory cells with
types and break when an inconsistency is detected (e.g., when a \texttt{double} is stored in a
cell pointed by a \texttt{int*} pointer).

\hone{Heavyweight Verification}
}

\myparag{Static Analysis and Abstract Interpretation}
With heavyweight verification techniques~\cite{DBLP:conf/popl/CousotC77},
the source code of the software is analyzed without being run.
The goal is to find errors and chunks of code that can cause maintenance difficulties that raise the risk of introducing bugs during subsequent modifications.
Properties can also be proven over the behavior of the software.
Unfortunately, theses approaches can be slow, limited to certain classes of bugs or safety properties and can produce false positives and false negatives.
SLAM is based on static analysis and aims at checking good API usage.
SLAM is restricted to system code, mainly Windows~\cite{Ball:2006:TSA:1217935.1217943,Ball:2002:SLP:503272.503274} drivers.

\hideInShortVersion{
\myparag{Model-Checking}
Model checking is an automatic verification technique for finite-state reactive systems.
Model checking consists in checking that a model of the system verifies temporal properties~\cite{DBLP:books/daglib/0007403}.

While static analysis and abstract interpretation and model checking can
provide certain guarantees by proving properties over the program or a model of the program, proving
correction of a software statically is undecidable in general~\cite{landi1992undecidability}.
}
\hideInShortVersion{
\hone{Monitoring}

Monitoring consists in property checking at runtime. Checks are performed on event
produced during the execution. Production of events requires instrumentation.
Different instrumentation techniques exist. In this section, we give some of
the most important ones.
}
\hideInShortVersion{
\myparag{Compile-Time Instrumentation}
RiTHM~\cite{DBLP:conf/sigsoft/NavabpourJWBMBF13} is an implementation of a
time-triggered monitor, i.e. a monitor ensuring predictable and evenly
distributed monitoring overhead by handling monitoring at predictable moments.
Instrumentation is added to the code of the program to monitor.
In our approach, the code of the program is not modified and not recompilation is required.
}

\myparag{Dynamic Binary Instrumentation}
DBI makes it possible to detect cache-related performance and memory usage related problems.
The monitored program is instrumented by dynamically adding instructions to its binary code at runtime and run in an virtual machine-like environment.
Valgrind~\cite{DBLP:conf/pldi/NethercoteS07} is a tool that leverages DBI and can interface with GDB.
It provides a way to detect memory-related defects.
\hideInShortVersion{
Dr. Memory~\cite{DBLP:conf/cgo/BrueningZ11} is another similar tool based on
DynamoRIO~\cite{DBLP:conf/vee/BrueningZA12}.
DynamioRIO and Intel Pin~\cite{DBLP:conf/pldi/LukCMPKLWRH05} are both DBI
frameworks that allow to write dynamic instrumentation-based analysis tools.
}
DBI provides a more comprehensive detection of memory-related defects than using the instrumentation tools provided by the debugger.
However, it is also less efficient and implies greater overheads when looking for particular defects like memory leaks caused by the lack of a call to the function~\texttt{free}.

\myparag{Instrumentation Based on the VM of the Language}
For some languages like Java, the Virtual Machine provides introspection and features like
aspects~\cite{DBLP:conf/jit/Kiczales02,DBLP:journals/taosd/Katz06} used to capture events.\hideInShortVersion{
The Jassda framework~\cite{brorkensdynamic}, which uses CSP-like specifications,
LARVA~\cite{CPS09larva,DBLP:conf/rv/ColomboFG11} and JavaMOP~\cite{chen2007mop}
are monitoring tools for virtual machine based programming languages~(mainly
Java).
}
This is different from our model which rather depends on the features of the debugger.
JavaMOP~\cite{DBLP:conf/icse/JinMLR12} is a tool that allows monitoring Java programs.
However, it is not designed for inspecting their internal state.
JavaMOP also implements trace slicing as described in~\cite{DBLP:conf/tacas/ChenR09}.
In our work, events are dispatched in slices in a similar way,
We do not implement all the concepts defined by~\cite{DBLP:conf/tacas/ChenR09} but this is sufficient for our purpose.
In monitoring, the execution of the program can also be affected by modifying the sequence of input or output events to make it comply with some properties~\cite{DBLP:conf/rv/Falcone10}.
This is different from our approach which applies earlier in the development cycle.
In our approach, we modify the execution from the inside and aim at fixing the program rather than its observable behavior.

\myparag{Debugger-Based Instrumentation}
Morphine, Opium and Co\-ca~\cite{ducasse2001efficient} are three automated trace analyzers.
The analyzed program is run in another process than the monitor, like in our approach.
The monitor is connected to a tracer.
Like in our approach, this work relies on the debugger to generate events.
Focus is set on trace analysis: interactivity is not targeted and the execution remains unaffected.

\myparag{Frama-C~\cite{DBLP:conf/sefm/CuoqKKPSY12}}
Frama-C is a modular platform aiming at analyzing source code written in C and provides plugins for static analysis, abstract interpretation, deductive verification, testing and monitoring programs.
It is a comprehensive platform for verification. It does not support interactive debugging nor programs written in other programming languages.



\myparag{Conclusions}
Current approaches to finding and studying bugs have their own drawbacks and benefits and are suitable for discovering different sorts of bugs in different situations.
Their relevance is also related to a phase of the program life cycle.
None of them gather bug discovery and understanding.

\ifdefined \SHORTVERSION

\hzero{Future Work}
\label{chap:future}

\else

\hzero{Conclusion and Future Work}
\label{chap:future}

\hone{Conclusion}
This report presents an approach combining monitoring and debugging as two
complementary approaches to program correctness.

In monitoring, the program receives and outputs events. Properties on these
events are verified or enforced. Detecting a bug is possible: if a
property on the correctness of the program breaks at runtime, a bug is present
in the program. However, a limitation of monitoring is that it does not provide
a way to understand bugs.

In debugging, the program has an internal state that
can be studied and modified. Interactive debugging is a way to understand a bug
and find its cause. However, debugging has no support for bug discovery: a
programmer uses a debugger at a time when the bug is already known.

Our approach aims at taking the best of both techniques by seeing the program
as a system that can be monitored to find bugs and, at the same time, as a
system that can be debugged interactively to understand the bugs that were
found. When a bug is found using monitoring, the debugger can be used in a
traditional way to understand it.

In this report, we described this approach in details, we provided a theoretical
framework that eases the reasoning about the notion of joint execution
of the monitor, the debugger and the program.
We also presented \ce{}, an implementation of this approach and its evaluation.
Our experiments showed that even though the property checker
can slow down the execution of the program considerably when events are
temporally close to each other, performance are acceptable beyond a reasonable
threshold (\secref{secmicrobench}).
We demonstrated that this approach is applicable in realistic use-cases with software such as video games and video
players~\ref{sec:video}.
Our current implementation shows limitations in terms of performance under other use-cases.
In the next section, we present ideas to mitigate this issue.
%
\hone{Future Work}
%
\fi

\ifdefined\RLYSHORTVERSION \else
In this section, we present some perspectives opened by this work: diversifying the supported event types, exploring other ways of instrumenting the execution, possibilities opened by checkpoints and further validating our approach.
\fi

\hideInShortVersion{
\myparag{Event Types}
Our main event types are the function call and variable accesses.
A way to make our approach more powerful is to find and include other kinds of events in our model.
System calls are an example of event type we have not taken in account yet for technical reasons.
They might be of interest for checking properties on drivers or programs dealing with hardware.
}

\myparag{Instrumentation}
Handling breakpoints is costly~\cite{chabot2015automatic} and handling watchpoints even more.
Code injection could provide better efficiency~\cite{DBLP:conf/sigsoft/NavabpourJWBMBF13,DBLP:journals/cl/MilewiczVTQP16}
by limiting round trips between the debugger and the program would to the bare
minimum\hideInShortVersion{ (for example, when the scenario requires the execution to be suspended
to let the user interact with the debugger)} while keeping the current
flexibility of the approach.
%


\myparag{Checkpointing the File System}
We plan to explore the possibility of capturing the environment of the developer in addition to the process being debugged when checkpointing.
More specifically, we shall look at the atomic snapshotting capabilities of modern file systems like Brfs and ZFS.

\myparag{Record and Replay and Reverse Execution}
RR is a powerful technique for finding bugs. Once a buggy
execution is recorded, the bug can be studied and observed again by running the
recording.
We aim to augment i-RV with reverse debugging and RR techniques.
%
%
%

\myparag{Validation of the Approach}
Our approach has been evaluated on small, simple examples.
Next step is to validate it in more concrete situations, find more cases of real bugs in
widespread applications and show that it indeed eases both discovery and understanding of the bug with a solid user study.

\hideInShortVersion{
Another idea that is yet to be explored is verifying good practice rules
and good API usage at runtime. We think that API designers and library writers
could leverage our approach by providing properties with their APIs and their
libraries.
This would provide a means to check that their APIs are used
correctly and make their usage safer. This would also be a means to document
these APIs and these libraries.
}

\bibliography{bibliography}
%
%
\ifdefined \SHORTVERSION \else
\appendix
\appendix

\hzero{\ce{} Command List}
\label{apdx:cecmdlist}
\begin{tabularx}{\linewidth}{XX}
\texttt{verde activate}   &  Activates all the commands monitor related commands \\
\texttt{verde checkpoint} &  Sets a checkpoint for the program and each managed monitor\\
\texttt{verde checkpoint-restart} &  Restores a checkpoint\\
\texttt{verde cmd-group-begin} &  Begins a group of commands\\
\texttt{verde cmd-group-end} &  Ends a group of command\\
\texttt{verde delete} &  Deletes a monitor\\
\texttt{verde exec} &  Executes an action in the current monitor. Can be used to call methods of the interface of the current monitor defined in Appendix~\ref{apdx:cemi}\\
\texttt{verde get-current} &  Prints the name of the current monitor\\
\texttt{verde load-functions} &  Loads a user defined functions file\\
\texttt{verde load-property} &  Loads a property file and possibly a function file in the given monitor\\
\texttt{verde load-scenario} &  Loads a property file and possibly a function file in the given monitor\\
\texttt{verde new} &  Creates a monitor that will also become the current monitor\\
\texttt{verde run} &  Runs the monitor\\
\texttt{verde run-with-program} &  Running the monitor and the program at the same time\\
\texttt{verde set-current} &  Sets the current monitor\\
\texttt{verde show-graph} &  Shows the graph of the monitor in a window and animates it at runtime
\end{tabularx}
\ifdefined\TWOCOLS \else
\hzero{Interface of the \Ce{} Monitors}
\label{apdx:cemi}

The following is the documentation of the MonitorInterface class. Its methods
can be used programmatically, some of them can be used from the shell of GDB
using the \verb!verde exec! command. For instance,
\verb!verde exec get_current_states! prints the current states of the
current monitor.

{\small
\begin{lstlisting}
debugger_shell(self)
    Raises an exception making the monitor drop to the shell of the debugger.

get_current_states(self)
    Returns the set of the current states of the
    property.

get_env(self, s=None)
    Returns the environment dictionary of the property.

get_env_keys(self, s, as_iterator=True)
    Returns the keys of the environment of the property, as an iterator or a
    list, whether as_iterator is True or False, repectively.

get_env_value(self, s, key)
    Returns the value of the given variable in the environment of the property
    environment. Raises if the key is not present in the environment.

get_slice_bindings(self)

get_slices(self)

get_states(self)
    Returns the set of the states of the property.

print_monitor_state(self)
    Prints current state of the monitor.

register_event(self, event_type, callback)
    Register a callback for this event type.

    Possible events:
     - state_changed(new_states)
        new_states is the set of the new current states
     - transition_taken(transition, point)
        transition is the object representing the transition
        point is either "success" or "failure"
     - event_applied

    The event type is given by its name and the parameters passed to the
    callback is what is given in parenthesis.

set_current_state(self, state)
    Sets the current state of (the root slice of) the property.

set_env_dict(self, s, new_env)
    Sets the environment of the property.

set_env_value(self, s, key, value)
    Sets the value of the given variable in the environment of the property.

set_globals(self, g)
    Sets the dictionary in which functions will be found, if needed,
    when e.g. calling (un)register_event.

set_quiet(self, b='True')
    Sets the monitor quiet or not.

set_transition_debug_function(self, fun_name=None)
    Specifies a user's function to call whenever a monitored action not
    taken in account in the current states of the property is called.
    no argument means the default: no user function is called when it
    happens.

step_by_step(self, b='True')
    Set step by step monitor

stop_execution(self)
    Raises an exception making the execution of the monitor stop the
    execution of the program and the monitor.

unregister_event(self, event_type, callback)
    Unregister a callback for this event type.
    See also register_event.

Some commands are not accessible from verde exec: get_env_dict,
set_globals, get_env_keys, stop_execution, set_current_states, debugger_shell,
set_env_dict.
\end{lstlisting}
}

\fi

\hzero{Property on value changes of s in get\_comp\_string in \zsh{}}
\label{apdx:zshprop}

In this appendix, we present a property written in \ce{} property format that is used to find the cause of a segfault in \zsh{}; see Figures~\ref{fig:prop1} and~\ref{fig:prop2}.
In this property, we are in an accepting state while the state of \zsh{} seems consistent.
That is, no null pointer is going to be used. In state init, we track a call to the function get\_comp\_string. When the call happens, the state becomes in\_get\_cmp\_str\_init. In this state, several things can happen.
Destination states in the state in\_get\_cmp\_str\_init correspond to the different continuations we imagined as possible after this state by taking a quick look at the code.
We did not aim at exactly understanding the meaning of these different possibilities.
Rather, we aimed at seeking where the pointer was nulled in the code.

Note that for the sake of readability, we omit importing gdb with the command \lstinline[style=CheckExecStyle]{import gdb} in the failure handlers.

\begin{figure}[t]
\begin{lstlisting}[style=CheckExecStyle,deletekeywords={max}]
state init accepting {
    transition {
        before event get_comp_string()
        success in_get_cmp_str_init
    }
}
state in_get_cmp_str_init accepting {
    transition {
        after event write s(s) { return s != None and s != 0 }
        success { print("s = " + str(s)) } in_get_cmp_str_init_s_not_null
        failure { gdb.execute("backtrace") } in_get_cmp_str_init_s_null
    }
    transition {
        before event itype_end(ptr) { return ptr == None or ptr == 0 }
        success calling_itype_end_with_null_ptr
    }
    transition {
        before event get_comp_string()
        success in_get_cmp_str_init
    }
}
state in_get_cmp_str_init_s_null accepting {
    transition {
        after event write s(s) { return s != None and s != 0 }
        success { print("s = " + str(s)) } in_get_cmp_str_init_s_not_null
        failure { gdb.execute("backtrace") } in_get_cmp_str_init_s_null
    }
    transition {
        before event itype_end(ptr) { return ptr == None or ptr == 0 }
        success calling_itype_end_with_null_ptr
        failure {
        	print("called itype_end with non null");
        	gdb.execute("backtrace")
        } in_get_cmp_str_init_s_null
    }
    transition {
        before event get_comp_string()
        success in_get_cmp_str_init
    }
}
...
\end{lstlisting}
\caption{\Ce{} property to find the cause of the segfault in \zsh{} - Part 1)}
\label{fig:prop1}
\end{figure}

\begin{figure}[t]
\begin{lstlisting}[style=CheckExecStyle,deletekeywords={max}]
state in_get_cmp_str_init_s_not_null accepting {
    transition {
        after event write s(s) { return s != None and s != 0 }
        success { print("s = " + str(s)) } in_get_cmp_str_init_s_not_null
        failure { gdb.execute("backtrace") } in_get_cmp_str_init_s_null
    }
    transition {
        before event itype_end(ptr) { return ptr == None or ptr == 0 }
        success calling_itype_end_with_null_ptr
    }
    transition {
        before event get_comp_string()
        success in_get_cmp_str_init
    }
}
state in_get_cmp_str_init_s_not_null accepting {
    transition {
        before event get_comp_string()
        success in_get_cmp_str_init
    }
    transition {
        after event write s(s) { return s != None and s != 0 }
        success { print("s = " + str(s)) } in_get_cmp_str_init_s_not_null
        failure { gdb.execute("backtrace") } in_get_cmp_str_init_s_null
    }
}
state calling_itype_end_with_null_ptr non-accepting { }
\end{lstlisting}
\caption{\Ce{} property to find the cause of the segfault in \zsh{} - Part 2)}
\label{fig:prop2}
\end{figure}

\fi

%
%
\end{document}